\documentclass[aps,a4paper,superscriptaddress,showpacs,preprintnumbers,amsmath,amssymb]{revtex4}

\usepackage{ulem}

\usepackage{psfrag} \usepackage{graphicx} \usepackage{dcolumn}
\usepackage{color} \usepackage{latexsym,amsfonts} \usepackage{bm}
\usepackage{amssymb} 
 
\begin{document}

\title{Neutron Dark Matter Decays}

\author{A. N. Ivanov}\email{ivanov@kph.tuwien.ac.at}
\affiliation{Atominstitut, Technische Universit\"at Wien, Stadionallee
  2, A-1020 Wien, Austria}
\author{R.~H\"ollwieser}\email{roman.hoellwieser@gmail.com}
\affiliation{Atominstitut, Technische Universit\"at Wien, Stadionallee
  2, A-1020 Wien, Austria}\affiliation{Department of Physics,
  Bergische Universit\"at Wuppertal, Gaussstr. 20, D-42119 Wuppertal,
  Germany} \author{N. I. Troitskaya}\email{natroitskaya@yandex.ru}
\affiliation{Atominstitut, Technische Universit\"at Wien, Stadionallee
  2, A-1020 Wien, Austria}
\author{M. Wellenzohn}\email{max.wellenzohn@gmail.com}
\affiliation{Atominstitut, Technische Universit\"at Wien, Stadionallee
  2, A-1020 Wien, Austria} \affiliation{FH Campus Wien, University of
  Applied Sciences, Favoritenstra\ss e 226, 1100 Wien, Austria}
\author{Ya. A. Berdnikov}\email{berdnikov@spbstu.ru}\affiliation{Peter
  the Great St. Petersburg Polytechnic University, Polytechnicheskaya
  29, 195251, Russian Federation}

\date{\today}

\begin{abstract}
We analyse the discrepancy between the neutron lifetimes measured in
the bottle and beam experiments. Following Fornal and Grinstein
(Phys. Rev. Lett. {\bf 120}, 191801 (2018)) we propose an explanation
of such a puzzle by the dark matter modes of the neutron
decay. However, unlike Fornal and Grinstein in addition to the dark
matter decay mode $n \to \chi + e^- + e^+$, where $\chi$ is a dark
matter Dirac fermion and $(e^-e^+)$ is an electron--positron pair, we
assume the existence of the dark matter mode $n \to \chi + \nu_e +
\bar{\nu}_e$, where $\nu_e \bar{\nu}_e$ is the neutrino--antineutrino
pair. This allows to describe the discrepancy between the measurements
of the neutron lifetime even in case of an unobservability of the dark
matter decay mode $n \to \chi + e^- + e^+$, which may be below the
reaction threshold. The existence of the coupling $n \to \chi + e^- +
e^+$ can be observed experimentally by measuring electron--neutron
scattering $e^- + n \to \chi + e^-$ at very low electron energies,
induced with the strength of the decay $n \to \chi + \nu_e +
\bar{\nu}_e$. We propose a gauge invariant quantum field theory model
with $SU_L(2)\times U_R(1) \times U_R'(1)\times U''_L(1)$ symmetry for
the UV completion of the effective $(n\chi \ell \bar{\ell})$
interaction, where $\ell(\bar{\ell})$ is electron (positron) or
neutrino(antineutrino). We show that predictions of our model do not
contradict to constraints on dark matter production in ATLAS
experiments at the LHC and to constraints reported by other
experimental groups, and evolution of neutron stars.  We argue that
reactions $n \to \chi + \nu_e + \bar{\nu}_e$, $n + n \to \chi + \chi$,
$n + n \to \chi + \chi + \nu_e + \bar{\nu}_e$ and $\chi + \chi \to n +
n$, allowed in our model, can serve as URCA processes for the neutron
star cooling. We argue that existence of neutron dark matter decays
demands the revision of the neutron $\beta^-$--decay beyond the
Standard Model with Fierz interference term of order $b \sim -
10^{-2}$. 
\end{abstract} 
\pacs{ 11.10.Ef, 13.30a, 95.35.+d, 25.40.Fq}

\maketitle

\section{Introduction}
\label{sec:introduction}

The neutron plays a dominant role in the fields of particle, nuclear
and neutrino physics, astrophysics and cosmology
\cite{Dubbers2011}. Today the measured neutron lifetime is still the
only source to derive various semileptonic charged weak cross sections
needed in this fields \cite{Ivanov2013a}. The world averaged value of
the neutron lifetime is now $\tau_n = 880.1(1.0)\,{\rm s}$
\cite{PDG2018}, but the experimental data on the neutron lifetime
measurements of in--beam (flight) experiments differ from the data of
in--trap (bottle) experiments by up to one percent, and disagree with
the calculated value. Indeed, the analysis of the neutron
$\beta^-$--decay with a polarized neutron and unpolarized electron and
proton, which has been carried out in \cite{Ivanov2013} with the
account for the complete set of corrections of order $10^{-3}$,
calculated in the Standard Model (SM) and caused by the weak magnetism
and proton recoil to next--to--leading order in the nucleon mass
expansion and radiative corrections, showed that the neutron lifetime,
calculated at the axial constant $\lambda = - 1.2750(9)$
\cite{Abele2008} (see also \cite{Abele2013}), is equal to $\tau^{(\rm
  SM)}_n = 879.6(1.1)\,{\rm s}$. The neutron lifetime $\tau^{(\rm
  bottle)}_n = 879.6(6)\,{\rm s}$, averaged over the experimental
values of the six bottle experiments
\cite{Mampe1993}--\cite{Arzumanov2015} included in the Particle Date
Group (PDG) \cite{PDG2018} (see also \cite{Czarnecki2018}), agrees
perfectly well with the value $\tau^{(\rm SM)}_n = 879.6(1.1)\,{\rm
  s}$. Recent experimental values of the neutron lifetime $\tau_n =
877.7^{+0.8}_{-0.7}\,{\rm s}$ \cite{Pattie2017} and $\tau_n =
881.5(9)\,{\rm s}$ \cite{Serebrov2017} agree with the averaged value
of the neutron lifetime $\tau^{(\rm bottle)}_n = 879.6(6)\,{\rm
  s}$. Thus, one may see that there is no room within the SM to get
the neutron lifetime larger than $\tau^{(\rm SM)}_n = 879.6(1.1)\,{\rm
  s}$ supported by the bottle experiments $\tau^{(\rm bottle)}_n =
879.6(6)\,{\rm s}$ (see also \cite{Czarnecki2018}).

Nevertheless, the measurements of the neutron lifetime in the beam
experiments show larger values \cite{Byrne1990}--\cite{Bowman2014}.
Indeed, the neutron lifetime $\tau^{(\rm beam)}_n = 888.0(2.0)\,{\rm
  s}$, averaged over the experimental values of the beam measurements
\cite{Byrne1990}--\cite{Bowman2014}, differs from the averaged one
$\tau^{(\rm bottle)}_n = 879.6(6)\,{\rm s}$, obtained from the bottle
experiments \cite{Mampe1993}--\cite{Arzumanov2015}, by $ \Delta \tau_n
= 8.4(2.1)\,{\rm s}$. For an explanation of this discrepancy Fornal
and Grinstein \cite{Fornal2018} have proposed that the total width
$\lambda_n$ of the neutron decay is defined by
\begin{eqnarray}\label{eq:1}
\lambda_n = \lambda_{n \to p} + \lambda_{n \to \chi},
\end{eqnarray}
where $\lambda_{n\to p}$ and $\lambda_{n \to \chi}$ are the widths,
caused by the decay modes $n \to p + {\rm anything}$ and $n \to
\chi + {\rm anything}$, respectively, and $\chi$ is a dark matter
Dirac particle. According to Fornal and Grinstein \cite{Fornal2018},
the dark matter decay modes can be $n \to \chi + \gamma$, $n \to \chi
+ \gamma^* \to \chi + e^- + e^+$, and $n \to \chi + \phi$, where
$\gamma$ and $\gamma^*$ are real and virtual photons,
  respectively, and $\phi$ is a scalar dark matter particle. The
total neutron lifetime equal to $\tau_n = 1/\lambda_n$ is measured in
the bottle experiments, $\tau_n = \tau^{(\rm bottle)}_n =
879.6(6)\,{\rm s}$.  In turn, in the beam experiments, where the decay
proton is detected, the neutron lifetime $\tau^{(\rm beam)}_n =
888.0(2.0)\,{\rm s}$ is defined by (see \cite{Fornal2018})
\begin{eqnarray}\label{eq:2}
\tau^{(\rm beam)}_n = \frac{\tau_n}{{\rm BR}(n \to p + {\rm
    anything})},
\end{eqnarray}
where ${\rm BR}(n \to p + {\rm anything}) = 0.99\,\%$ in order to
explain the neutron lifetime puzzle in such a way. Since the neutron
lifetime $\tau_n = 879.6(1.1)\,{\rm s}$, calculated in the SM with the
axial coupling constant $\lambda = - 1.2750(9)$ the value of which is
supported by a global analysis by Czarnecki {\it et al.}
\cite{Czarnecki2018}, agrees perfectly well with the neutron lifetime
$\tau_n = 879.6(6)\,{\rm s}$, averaged over the experimental data of
bottle experiments, the possibility for the neutron to have any dark
matter decay mode is fully ruled out. Indeed, according to the
hypothesis by Fornal and Grinstein \cite{Fornal2018}, the SM should
explain the neutron lifetime $\tau_n = 888.0(2.0)\,{\rm s}$, measured
in the beam experiments, instead of to explain the neutron lifetime
$\tau_n = 879.6(6)\,{\rm s}$, measured in bottle ones. In order to fit
the value $\tau_n = 888.0(2.0)\,{\rm s}$ by the analytical expression
for the neutron lifetime (see Eq.(41) and (42) of
Ref.\cite{Ivanov2013}) the axial coupling constant $\lambda$ should be
equal to $\lambda = - 1.2690$. Since such a value of the axial
coupling constant is ruled out by recent experiments
\cite{Abele2013}--\cite{Brown2018} and global analysis by Czarnecki
     {\it et al.} \cite{Czarnecki2018}, so the hypothesis by Fornal
     and Grinstein \cite{Fornal2018} should state that the SM,
     including a complete set of corrections of order $10^{-3}$ caused
     by the weak magnetism, proton recoil and radiative corrections
     \cite{Gudkov2006} (see also \cite{Ivanov2013}), is not able to
     describe correctly the neutron decay modes $n \to p + {\rm
       anything}$. Hence, the theoretical description of the neutron
     lifetime, measured in the beam experiments, should go beyond the
     SM. Indeed, keeping the value of the axial coupling constant
     equal $\lambda = - 1.2750$ or so
     \cite{Abele2013}--\cite{Brown2018} and having accepted an
     existence of the dark matter decay modes $n \to \chi + {\rm
       anything}$ we have also to accept a sufficiently large
     contribution of the Fierz interference term $b$
     \cite{Rose1955}. Using the results obtained in \cite{Ivanov2013},
     the neutron lifetime $\tau_n = 888.0\,{\rm s}$ can be fitted by
     the axial coupling constant $\lambda = -1.2750$, the
     Cabibbo--Kobayashi--Maskawa (CKM) matrix element $V_{ud} =
     0.97420$ \cite{PDG2018} and the Fierz interference term $b = -
     1.44\times 10^{-2}$ \cite{Ivanov2018g}.  This is the price for
     the acceptance of the neutron dark matter decay modes, explaining
     the neutron lifetime anomaly.  Keeping in mind such an important
     problem, which has at least one positive solution
     \cite{Ivanov2018g}, of the neutron $\beta^-$--decays, caused by
     the acceptance of the neutron dark matter decays $n \to \chi +
     anything$, we may proceed to the discussion of the properties of
     dark matter fermions from the neutron dark matter decay modes $n
     \to \chi + {\rm anything}$ and their compatibility with different
     experimental data by the ATLAS Collaboration at the LHC, dynamics
     of neutron stars and so on.

According to recent experimental data \cite{Tang2018}, the decay mode
$n \to \chi + \gamma$ is suppressed. This entails a suppression of the
decay mode $n \to \chi + \gamma^* \to \chi + e^- + e^+$. Then, in the
experiment by the UCNA Collaboration \cite{Sun2018} there has been
found that the decay $n \to \chi + e^- + e^+$ is excluded as a
dominant dark matter decay mode at the level of $\gg 5\sigma$ for the
kinetic energies $T_{-+}$ of the electron--positron pairs constrained
by $100\,{\rm keV} < T_{-+} < 644\,{\rm keV}$. As has been pointed out
in \cite{Sun2018}, if the final state $\chi + e^- + e^+$ is not only
one, the limit on its branching fraction is $< 10^{-4}$ for $100\,{\rm
  keV} < T_{-+} < 644\,{\rm keV}$ at $> 90\,\%$ (C.L.).  The latter
does not contradict our assumption about an existence of the dark
matter decay mode $n \to \chi + \nu_e + \bar{\nu}_e$.  Thus, according
to the experimental data by \cite{Tang2018,Sun2018}, an existence of
the decay mode $n \to \chi + e^- + e^+$ seems to be
suppressed. Of course, there is room for the following
  assumptions: i) the dark matter decay mode $n \to \chi + e^- + e^+$
  is not mediated by a virtual photon $\gamma^*$ but induced an
  effective phenomenological low--energy interaction, ii) the
  electron--positron pair in the decay mode $n \to \chi + e^- + e^+$,
  produced by such an interaction, can be detected either by the UCNA
  Collaboration \cite{Sun2018} for kinetic energies of the
  electron--positron pair $T_{-+} < 100\,{\rm keV}$ or by the PERKEO
  Collaboration \cite{Klopf2018} using the electron spectrometer
  PERKEO II \cite{Abele2016}, and iii) an unobservability of the
  production of the electron--positron pair from the decay mode $n \to
  \chi + e^- + e^+$, induced by an effective phenomenological
  low--energy interaction, may only mean that the electron--positron
  pair production in the decay $n \to \chi + e^- + e^+$ is below the
  reaction threshold, i.e. $m_n < m_{\chi} + 2m_e$, where $m_n$,
  $m_{\chi}$ and $m_e$ are the masses of the neutron, dark matter
  fermion and electron (positron), respectively.

This paper is addressed to the theoretical analysis of the neutron
dark matter decay modes $n \to \chi + e^- + e^+$ and $n \to \chi +
\nu_e + \bar{\nu}_e$.  For the extended analysis of the neutron dark
matter decay modes we may in principle use the following most general
phenomenological Lagrangian of dark--matter-baryon-lepton interactions
\begin{eqnarray}\label{eq:3}
\hspace{-0.3in}&&{\cal L}_{\rm DMBL}(x) = -
\frac{G_F}{\sqrt{2}}\,V_{ud}\,\Big\{[\bar{\psi}_{\chi}(x)\gamma_{\mu}
  \psi_n(x)] [\bar{\Psi}_e(x)\gamma^{\mu}(h_V + \bar{h}_V
  \gamma^5)\Psi_e(x)]\nonumber\\
\hspace{-0.3in}&& +
       [\bar{\psi}_{\chi}(x)\gamma_{\mu}\gamma^5\psi_n(x)]
       [\bar{\Psi}_e(x) \gamma^{\mu}(\bar{h}_A + h_A
         \gamma^5)\Psi_e(x)] +
       [\bar{\psi}_{\chi}(x)\psi_n(x)][\bar{\Psi}_e(x)(h_S + \bar{h}_S
         \gamma^5)\Psi_e(x)]\nonumber\\
\hspace{-0.3in}&& + [\bar{\psi}_{\chi}(x) \gamma ^5
  \psi_n(x)][\bar{\Psi}_e(x)(h_P + \bar{h}_P \gamma^5)\Psi_e(x)] +
\frac{1}{2} [\bar{\psi}_{\chi}(x)\sigma^{\mu\nu} \gamma^5\psi_n(x)]
     [\bar{\Psi}_e(x)\sigma_{\mu\nu} (\bar{h}_T + h_T
       \gamma^5)\Psi_e(x) \Big\},
\end{eqnarray}
where $G_F = 1.1664\times 10^{-11}\,{\rm MeV}^{-2}$ is the Fermi weak
constant, $V_{ud} = 0.97420(21)$ is the Cabibbo-Kobayashi--Maskawa
(CKM) matrix element \cite{PDG2018}, extracted from the $0^+ \to 0^+$
transitions \cite{PDG2018}. The factor $G_FV_{ud}/\sqrt{2}$ is
introduced for convenuence to compare the neutron dark matter decays
$n \to \chi + e^- + e^+$ and $n \to \chi + \nu_e + \bar{\nu}_e$ with
the neutron decays $n \to p + {\rm anything}$. The coupling constants
$h_j$ and $\bar{h}_j$ for $j = V,A,S,P$ and $T$ are phenomenological
coupling constants of vector, axial--vector, scalar, pseudoscalar and
tensor interactions, respectively, define the strength of the
dark--matter--neutron--lepton interactions. The Lagrangian
Eq.(\ref{eq:3}) is obviously relativistic covariant or
  invariant under Lorentz transformations \cite{Itzykson1980} and
written by analogy with well--known phenomenological
interactions beyond the SM, proposed in
\cite{Rose1955,Lee1956}--\cite{Severijns2006} (see also
\cite{Ivanov2013,Ivanov2018a}--\cite{Gardner2013}).  Then,
$\psi_{\chi}(x)$ and $\psi_n(x)$ are the field operators of the dark
matter Dirac particle and neutron, respectively. According to the
Standard Electroweak Model (SEM) \cite{PDG2018}, the field operator
$\Psi_e(x)$ is defined by
\begin{eqnarray}\label{eq:4}
\Psi_e(x) = \left(\begin{array}{c}\psi_{\nu_e}(x)
    \\ \psi_e(x)
\end{array}\right),
\end{eqnarray}
where $\psi_{\nu_e}(x)$ and $\psi_e(x)$ are the field operators of the
electron--neutrino (electron--antineutrino) and electron (positron),
respectively. If we take into account that in the SEM
  the neutrino--electron doublet is left--handed, i.e. $\Psi_e(x) \to
  \Psi_{eL}(x) = P_L\Psi_e(x)$ with the projection operator $P_L = (1
  - \gamma^5)/2$, the effective phenomenological interaction
  Eq.(\ref{eq:3}) reduces to the form
\begin{eqnarray}\label{eq:5}
{\cal L}_{\rm DMBL}(x) = -
\frac{G_F}{\sqrt{2}}\,V_{ud}[\bar{\psi}_{\chi}(x)\gamma_{\mu}(h_V +
  \bar{h}_A\gamma^5)\psi_n(x)] [\bar{\Psi}_e(x)\gamma^{\mu}(1 -
  \gamma^5)\Psi_e(x)],
\end{eqnarray}
where without loss of generality we have set $\bar{h}_j = - h_j$ for
$j = V,A$.  Using the Lagrangian Eq.(\ref{eq:5}) we calculate i) the
electron--energy and angular distribution of the decay $n \to \chi +
e^- + e^+$ mode, ii) the contribution of the dark matter decay mode $n
\to \chi + e^- + e^+$ to probability distribution of the neutron
$\beta^-$--decays as a function of the electron energy, iii) the
probability distribution of the neutron decay mode $n \to \chi + e^- +
e^+$ as a function of the electron energy, iv) the energy and angular
distribution and the rate of the decay mode $n \to \chi + \nu_e +
\bar{\nu}_e$, v) the probability distribution of the decay $n \to \chi
+ e^- + e^+$ as a function of the invariant mass of the
electron--positron pair and vi) the differential cross section for the
reaction $e^- + n \to \chi + e^-$ of the low--energy electron
scattering by polarized and unpolarized neutron. The results, obtained
in i) - v) can be used as a theoretical background for searches of the
dark matter decay mode $n \to \chi + e^- + e^+$ in the region of
kinetic energies $T_{-+} < 100\,{\rm keV}$ of the electron--positron
pairs in experiments of the UCNA Collaboration \cite{Sun2018} and of
the PERKEO II Collaboration \cite{Klopf2018}.

The paper is organized as follows. In section \ref{sec:betaspestrum}
we calculate the electron--energy and angular distribution of the dark
matter decay $n \to \chi + e^- + e^+$ for polarized neutron and
unpolarized decay particles, and the rate of this decay mode. We give
also the probability distribution of the neutron decays as a function
of the electron energy. In section \ref{sec:probability} we give i)
the theoretical probability distribution of the dark matter decay $n
\to \chi + e^- + e^+$ as a function of the electron energy and the
correlation coefficient of the electron--positron 3--momenta. In
section \ref{sec:nuspestrum} we calculate the
electron--antineutrino--energy and angular distribution of the dark
matter decay $n \to \chi + \nu_e + \bar{\nu}_e$ and the contribution
of this decay mode to the rate of the neutron decays. The numerical
value of the correlation coefficient $\zeta^{(\rm dm)}$, defining the
strength of the phenomenological $n \chi e^- e^+$ and $n \chi \nu_e
\bar{\nu}_e$ couplings, is estimated at the assumption that the
production of the electron--positron pair in the dark matter decay $n
\to \chi + e^- + e^+$ is below the reaction threshold, i.e. the mass
of the dark matter fermion obeys the constraint $m_{\chi} > m_n - 2
m_e$. In section \ref{sec:asymm} we give the theoretical expression
for the electron asymmetry, caused by the electron 3--momentum and
neutron spin correlations. In section \ref{sec:mass} we give the
probability distribution of the decay $n \to \chi + e^- + e^+$
relative to the rate of the neutron $\beta^-$--decay as a function of
the invariant mass of the electron--positron pair.  In section
\ref{sec:crosssection} we calculate the differential cross section and
the cross section for the low--energy electron--neutron scattering
$e^- + n \to \chi + e^-$. The differential cross section possesses the
following properties: i) it is inversely proportional to velocity of
incoming electrons, ii) it is isotropic relative to directions of the
3--momentum of outgoing electrons $\vec{k}^{\,'}_e$, and iii) the
absolute values of the 3--momenta of outgoing electrons are much
larger than the momenta of incoming electrons. All of these properties
of the differential cross section for the reaction $e^- + n \to \chi +
e^-$ allow to distinguish such a reaction above the background of the
elastic low--energy electron--neutron scattering.  In section
\ref{sec:uvcompletion} we i) propose a gauge invariant quantum field
theory model with $SU_L(2) \times U_R(1) \times U'_R(1)\times
U''_L(1)$ gauge symmetry for the UV completion of the effective
interaction Eq.(\ref{eq:5}), ii) discuss compatibility of predictions
of our model with constraints on the dark matter production in ATLAS
experiments at the LHC and on cross section for low--energy dark
matter fermion--electron scattering, iii) analyse an influence of dark
matter fermions with masses $m_{\chi} < m_n$ on dynamics of neutron
stars and iv) argue that processes $n \to \chi + \nu_e + \bar{\nu}_e$,
$n + n \to \chi + \chi$, $n + n \to \chi + \chi + \nu_e + \bar{\nu}_e$
and $\chi + \chi \to n + n$, allowed in our model, can serve as URCA
processes for the neutron star cooling.  In section
\ref{sec:conclusion} we discuss the obtained results and further
perspectives of our model for the analysis of dark matter in
terrestrial laboratories. We discuss also i) a possibility to avoid
violation of renormalizability of the quantum field theory of the dark
matter sector with $U'_R(1)$ gauge symmetry to any order
$O(e^n_{\chi}/2^n)$ of perturbation theory for $n \ge 6$
\cite{Bouchiat1972}--\cite{Bjorken1973}, caused by the
Adler--Bell--Jackiw anomaly \cite{Adler1969,Bell1969}, where
$e_{\chi}$ is the gauge coupling constant, and ii) constraints on the
mass of the dark matter spin--1 boson $Z'$ and the gauge coupling
constant $e_{\chi}$ from the branching ratio of the Higgs--boson decay
mode $H^0 \to Z + Z'$. We show that the branching ratio ${\rm Br}(H^0
\to Z Z') = 1.77\times 10^{-6}$, calculated in our model at the gauge
coupling constant $e_{\chi} = 1$, the mass $M_{Z'} \simeq 3\,{\rm
  GeV}$ of the dark matter spin--1 boson $Z'$, the mass $M_{H^0} =
125\,{\rm GeV}$ of the SM Higgs--boson \cite{ATLAS2012,CMS2012} and
its total width $\Gamma_{H^0} = 4.07\,{\rm MeV}$ \cite{Curtin2014},
agrees with the constraint ${\rm Br}(H^0 \to Z Z') \sim (10^{-4} -
10^{-6})$ imposed by Curtin {\it et al.}  \cite{Curtin2014} (see
Fig.\,12 of Ref. \cite{Curtin2014}). We propose also a version of our
model formulated at the quark level.

\section{Electron--energy and angular distribution of 
decay mode $n \to \chi + e^- + e^+$}
\label{sec:betaspestrum}

\begin{figure}
\includegraphics[height=0.15\textheight]{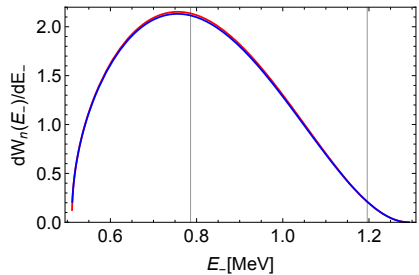}
  \caption{The probability distribution of the neutron decay modes
    Eq.(\ref{eq:9}) as a function of the electron energy and the mass
    $m_{\chi}$ of the dark matter fermion is plotted in the
    electron--energy region $m_e \le E_- \le E_0$ for $937.9\,{\rm
      MeV} \le m_{\chi} \le 938.543\,{\rm MeV}$, where the red and
    blue curves are defined for $m_{\chi} = 937.9\,{\rm MeV}$ and
    $m_{\chi} = 938.543\,{\rm MeV}$, respectively. }
\label{fig:fig1}
\end{figure}

Following \cite{Ivanov2013} the electron--energy and angular
distribution for the polarized neutron and unpolarized decay fermions
can be defined by
\begin{eqnarray}\label{eq:6}
\hspace{-0.3in}&&\frac{d^5 \lambda_{n \to
    \chi\,e^-e^+}(E_-,\vec{\xi}_n, \vec{k}_-, \vec{k}_+)}{dE_-
  d\Omega_-d\Omega_+} = (1 + 3
\lambda^2)\,\frac{G^2_F|V_{ud}|^2}{32\pi^5}\,({\cal E}^{(-+)}_0 + m_e
- E_-)\sqrt{({\cal E}^{(-+)}_0 - E_-)({\cal E}^{(-+)}_0 + 2m_e - E_-)}
\nonumber\\
\hspace{-0.3in}&&\times\,\sqrt{E^2_- - m^2_e}\, E_-\,
\zeta^{(\rm dm)}\,\Big(1 + a^{(\rm dm)}\,\frac{\vec{k}_-\cdot
  \vec{k}_+}{E_-E_+} + A^{(\rm dm)}_-\,\frac{\vec{\xi}_n\cdot
  \vec{k}_-}{E_-} + A^{(\rm dm)}_+\,\frac{\vec{\xi}_n\cdot
  \vec{k}_+}{E_+} + D^{(\rm dm)}\,\frac{\vec{\xi}_n\cdot (\vec{k}_-
  \times \vec{k}_+)}{E_-E_+}\Big),
\end{eqnarray}
where $\vec{\xi}_n$ is a unit vector of the neutron polarization. The
correlation coefficients $\zeta^{(\rm dm)}$, $a^{(\rm dm)}$, $A^{(\rm
  dm)}_{\mp}$ are equal to
\begin{eqnarray}\label{eq:7}
\hspace{-0.3in}\zeta^{(\rm dm)} &=& \frac{1}{1 + 3\lambda^2}\,\big(
|h_V|^2 + 3 |\bar{h}_A|^2\big)\;,\; a^{(\rm dm)} = \frac{|h_V|^2 -
  |\bar{h}_A|^2}{|h_V|^2 + 3 |\bar{h}_A|^2},\nonumber\\
\hspace{-0.3in} A^{(\rm dm)}_{\mp} &=& - 2\,\frac{{\rm
    Re}(h_V\bar{h}^*_A) \pm |\bar{h}_A|^2}{|h_V|^2 + 3
  |\bar{h}_A|^2}\;,\; D^{(\rm dm)} = \frac{2\,{\rm
    Im}(h_V\bar{h}^*_A)}{|h_V|^2 + 3 |\bar{h}_A|^2}.
\end{eqnarray}
Then, $(E_{\mp}, \vec{k}_{\mp})$ are the energies and 3--momenta of
the electron and positron, respectively, $d\Omega_{\mp} =
\sin\theta_{\mp} d\theta_{\mp} d\phi_{\mp}$ are the infinitesimal
solid angles of the electron $(-)$ and positron $(+)$ 3--momenta and
$\vec{k}_-\cdot \vec{k}_+ = k_-k_+ (\cos \theta_- \cos\theta_+ +
\sin\theta_- \sin\theta_+ \cos(\phi_- - \phi_+))$.  Then, ${\cal
  E}^{(-+)}_0 = (m^2_n - m^2_{\chi} - 2 m_{\chi}m_e)/2 m_n = m_n -
m_{\chi} - m_e$ is the end--point energy of the electron--energy
spectrum, taken to leading order in the large dark matter fermion and
neutron mass expansion. The positron energy $E_+$ and momentum $k_+$
are equal to $E_+ = {\cal E}^{(-+)}_0 + m_e - E_-$ and $k_+ =
\sqrt{({\cal E}^{(-+)}_0 - E_-)({\cal E}^{(-+)}_0 + 2 m_e - E_-)}$.
The rate of the dark matter decay mode $n \to \chi + e^- + e^+$ is
given by
\begin{eqnarray}\label{eq:8}
\lambda_{n \to \chi\,e^-e^+} &=& (1 + 3 \lambda^2)\,\zeta^{(\rm
  dm)}\,\frac{G^2_F|V_{ud}|^2}{2\pi^3}\int^{{\cal
    E}^{(-+)}_0}_{m_e}dE_-\,({\cal E}^{(-+)}_0 + m_e -
E_-)\sqrt{({\cal E}^{(-+)}_0 - E_-)({\cal E}^{(-+)}_0 + 2m_e - E_-)}
\nonumber\\
\hspace{-0.3in}&&\times\,\sqrt{E^2_- - m^2_e}\, E_-.
\end{eqnarray}
For the probability distribution of the neutron decay modes as a
function of the electron energy we obtain the following expression.
\begin{eqnarray}\label{eq:9}
\hspace{-0.3in}&&\frac{dW_n(E_-)}{dE_- } = \frac{(E_0 -
  E_-)^2 \,\sqrt{E^2_- - m^2_e}\, E_-}{f_n}\,\zeta^{(\rm
  SM)}(E_-)\,F(E_-, Z = 1)\Bigg\{1 + \Theta({\cal E}^{(-+)}_0 -
E_-)\,\frac{9.68\times 10^{-3} f_n}{(E_0 -
  E_-)^2}\nonumber\\
\hspace{-0.3in}&&\times \frac{({\cal E}^{(-+)}_0 + m_e -
  E_-)\sqrt{({\cal E}^{(-+)}_0 - E_-)({\cal E}^{(-+)}_0 + 2m_e -
    E_-)}}{\displaystyle \int ^{{\cal E}^{(-+)}_0}_{m_e}({\cal
    E}^{(-+)}_0 + m_e - E_-)\sqrt{({\cal E}^{(-+)}_0 - E_-)({\cal
      E}^{(-+)}_0 + 2m_e - E_-)}\,\sqrt{E^2_- - m^2_e}\,
  E_-\,dE_-}\Bigg\},
\end{eqnarray}
where $\Theta({\cal E}^{(-+)}_0 - E_-)$ is the Heaviside function,
$E_0 = (m^2_n - m^2_p + m^2_e)/2 m_n = 1.2927\,{\rm MeV}$ is the
end--point energy of the electron--energy spectrum \cite{Ivanov2013},
$f_n = 0.0616\,{\rm MeV^5}$ is the phase--volume of the neutron
$\beta^-$--decay \cite{Ivanov2013}, $F(E_-, Z = 1)$ is the
relativistic Fermi function and $\zeta^{(\rm SM)}(E_-)$ is the
correlation coefficient, calculated in the SM \cite{Gudkov2006} (see
also \cite{Ivanov2013}). In Fig.\,\ref{fig:fig1} we plot
Eq.(\ref{eq:9}) as a function of $m_{\chi}$ in the region $937.9\,{\rm
  MeV} \le m_{\chi} \le 938.543\,{\rm MeV}$, proposed by Fornal and
Grinstein \cite{Fornal2018}.
One may see that the contribution of the dark matter decay $n \to \chi
+ e^- + e^+$ to the electron--energy distribution can be hardly
distinguished above the electron--energy distribution from the neutron
$\beta^-$--decay $n \to p + e^- + \bar{\nu}_e$. In order to get the
positron--energy and angular distribution of the dark matter decay $n
\to \chi + e^- + e^+$ one may use Eq.(\ref{eq:6}) with the replacement
$E_- \to E_+$.

\section{Electron--energy and angular distribution of probability of
 decay mode $n \to \chi + e^- + e^+$ relative to rate of neutron
 $\beta^-$--decay}
\label{sec:probability}

Since, as it is seen from Fig.\,\ref{fig:fig1}, the contribution of
the dark matter decay $n \to \chi + e^- + e^+$ can be hardly
distinguished above the electron--energy distribution of the neutron
$\beta^-$--decay $n \to p + e^- + \bar{\nu}_e$, for the experimental
analysis of the dark matter decay $n \to \chi + e^- + e^+$ we propose
to use the probability distribution of this decay as function of the
electron--energy and angular correlations of the electron--positron
pairs. The probability distribution of the dark matter decay $n \to
\chi + e^- + e^+$, calculated relative to the rate of the neutron
$\beta^-$--decay from Eq.(\ref{eq:6}) for unpolarized neutron and
decay fermions, is equal to
\begin{eqnarray}\label{eq:10}
\hspace{-0.3in}&&\frac{d^5 W_{n \to \chi\,e^-e^+}(E_-, \vec{k}_-,
  \vec{k}_+)}{dE_- d\Omega_-d\Omega_+} =\nonumber\\
\hspace{-0.3in}&&= \frac{1}{16\pi^2}\,\frac{9.68\times 10^{-3}({\cal
    E}^{(-+)}_0 + m_e - E_-)\sqrt{({\cal E}^{(-+)}_0 - E_-)({\cal
      E}^{(-+)}_0 + 2m_e - E_-)(E^2_- - m^2_e)}\,E_-}{\displaystyle
  \int ^{{\cal E}^{(-+)}_0}_{m_e}({\cal E}^{(-+)}_0 + m_e -
  E_-)\sqrt{({\cal E}^{(-+)}_0 - E_-)({\cal E}^{(-+)}_0 + 2m_e -
    E_-)}\,\sqrt{E^2_- - m^2_e}\, E_-\,dE_-} \nonumber\\
\hspace{-0.3in}&&\times\,\
\Big(1 + a^{(\rm dm)}\,\frac{\sqrt{(E^2_- -
    m^2_e)({\cal E}^{(-+)}_0 - E_-)({\cal E}^{(-+)}_0 + 2 m_e -
    E_-)}}{E_-({\cal E}^{(-+)}_0 + m_e - E_-)}\,\big(\cos \theta_-
\cos\theta_+ + \sin\theta_- \sin\theta_+ \cos(\phi_- - \phi_+)\big)\Big).
\end{eqnarray}
\begin{figure}
\includegraphics[height=0.20\textheight]{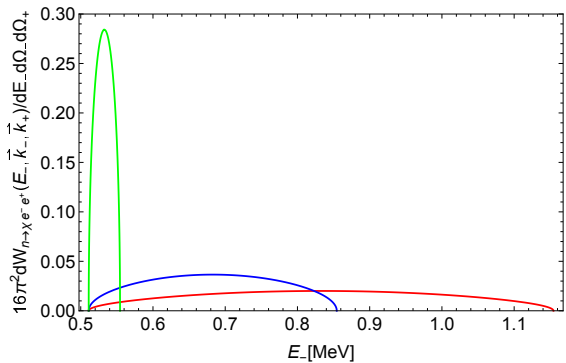}
  \caption{The probability distribution of the neutron dark matter
    decay $n \to \chi + e^- + e^+$ Eq.(\ref{eq:10}), calculated at
    $a^{(\rm dm)} = 0$, as a function of the electron energy and the
    mass $m_{\chi}$ of the dark matter fermion is plotted in the
    electron--energy region $m_e \le E_- \le {\cal E}^{(-+)}_0$ for
    $m_{\chi} = 937.9\,{\rm MeV}$ (red), $m_{\chi} = 938.2\,{\rm MeV}$
    (blue), and $m_{\chi} = 938.5\,{\rm MeV}$ (green), respectively. }
\label{fig:fig2}
\end{figure}
 The electron--energy probability distribution Eq.(\ref{eq:10}) can be
 used for the analysis of the experimental data on the search for the
 dark matter decay mode $n \to \chi + e^- + e^+$ in the region of
 electron--positron kinetic energies $T_{-+} < 100\,{\rm keV}$ by the
 UCNA Collaboration \cite{Sun2018} or by the PERKEO Collaboration
 \cite{Klopf2018} using the electron spectrometer PERKEO II. In order
 to get the positron--energy and angular distribution of the dark
 matter decay $n \to \chi + e^- + e^+$ one may use Eq.(\ref{eq:10})
 with the replacement $E_- \to E_+$. Unobservability of the
     electron--positron pairs in the neutron decays may mean that a
     production of such a pair is below threshold of the reaction $n
     \to \chi + e^- + e^+$. In case of unobservability of the dark
     matter decay mode $n \to \chi + e^- + e^+$ the rate of the
     neutron dark matter decay should be fully defined by the dark
     matter decay one $n \to \chi + \nu_e + \bar{\nu}_e$.

\section{Rate and antineutrino--energy and angular distribution of 
the neutron dark matter decay mode $n \to \chi + \nu_e + \bar{\nu}_e$
}
\label{sec:nuspestrum}

For the calculation of the rate of the dark matter decay mode $n \to
\chi + \nu_e + \bar{\nu}_e$ we have to calculate the
antineutrino(neutrino)--energy and angular distribution of this mode.
The antineutrino--energy and angular distribution of the neutron dark
matter decay mode $n \to \chi + \nu_e + \bar{\nu}_e$ for the
unpolarized massive fermions can be defined by
\begin{eqnarray}\label{eq:11}
\hspace{-0.3in}\frac{d^3 \lambda_{n \to \chi\,\nu_e
    \bar{\nu}_e}(E_{\bar{\nu}}, \vec{k}_{\bar{\nu}},
  \vec{k}_{\nu})}{dE_{\bar{\nu}} d\Omega} = (1 + 3
\lambda^2)\,\frac{G^2_F|V_{ud}|^2}{8\pi^4}\,({\cal E}_0 -
E_{\bar{\nu}})^2 \,E^2_{\bar{\nu}}\,\zeta^{(\rm dm)}\,\Big(1 + a^{(\rm
  dm)}\,\frac{\vec{k}_{\bar{\nu}}\cdot \vec{k}_{\nu}
}{E_{\bar{\nu}}E_{\nu}} \Big),
\end{eqnarray}
where the correlation coefficients $\zeta^{(\rm dm)}$ and $a^{(\rm
  dm)}$ are given in Eq.(\ref{eq:7}), and $d\Omega = \sin\vartheta
d\vartheta d\varphi$ is the infinitesimal solid angle of the
antineutrino--neutrino 3--momentum correlations
$\vec{k}_{\bar{\nu}}\cdot \vec{k}_{\nu} = E_{\bar{\nu}}E_{\nu}
\cos\vartheta$. The end--point energy of the antineutrino--energy
spectrum ${\cal E}_0$ is equal to ${\cal E}_0 = (m^2_n - m^2_{\chi})/2
m_n = m_n - m_{\chi}$ and taken to leading order in the large dark
matter fermion and neutron mass expansion. The rate of the decay mode
$n \to \chi + \nu_e + \bar{\nu}_e$ is equal to
\begin{eqnarray}\label{eq:12}
\lambda_{n \to \chi\,\nu_e \bar{\nu}_e} &=& (1 + 3
\lambda^2)\,\zeta^{(\rm
  dm)}\,\frac{G^2_F|V_{ud}|^2}{2\pi^3}\,\frac{{\cal E}^5_0}{30} =
\frac{\zeta^{(\rm dm)}}{\tau_n f_n}\,\frac{{\cal E}^5_0}{30} =
\frac{\tau^{(\rm beam)}_n - \tau^{(\rm bottle)}_n}{\tau^{(\rm
    beam)}_n\tau^{(\rm bottle)}_n} = 1.1 \times 10^{-5}\,{\rm s^{-1}},
\end{eqnarray}
where we have used $1/\tau_n = (1 +
  3\lambda^2)\,G^2_F|V_{ud}|^2f_n/2\pi^3$, $\tau_n = 879.6\,{\rm s}$
and $f_n = 0.0616\,{\rm MeV^5}$, which is the Fermi integral of the
neutron $\beta^-$--decay \cite{Ivanov2013}. Now we may define the
correlation coefficient $\zeta^{(\rm dm)}$
\begin{eqnarray}\label{eq:13}
\zeta^{(\rm dm)} = 1.1\times 10^{-5}\times 30\,\frac{\tau_nf_n}{{\cal
    E}^5_0} = \frac{0.018}{(m_n - m_{\chi})^5},
\end{eqnarray}
where $m_n - m_{\chi}$ is measured in MeV.  The correlation
coefficient $\zeta^{(\rm dm)}$ or the dimensionless coupling constant
Eq.(\ref{eq:13}) can be used for the analysis of the low--energy
electron--neutron inelastic $e^- + n \to \chi + e^-$
scattering. The estimate Eq.(\ref{eq:13}) is also
  valid if we replace $\tau_n = 879.6\,{\rm s}$ by $\tau_n =
  888.0\,{\rm s}$.

\section{Electron asymmetry of neutron decays with polarized neutron 
and unpolarized massive decay fermions}
\label{sec:asymm}

Keeping in mind that the UCNA and PERKEO
  Collaborations are able in principle to observe the dark matter
  decay mode $n \to \chi + e^- + e^+$ for electron--positron kinetic
  energies $T_{-+} < 100\,{\rm keV}$ we calculate the contribution of
  the decay mode $n \to \chi + e^- + e^+$ to the electron asymmetry of
  the neutron $\beta^-$--decay. Following \cite{Ivanov2013} we
calculate the contribution of the neutron dark matter decay $n \to
\chi + e^- + e^+$ to the electron asymmetry
\begin{eqnarray}\label{eq:14}
\hspace{-0.3in} A_{\exp} &=& \Big\{A^{(\rm SM)}(E_e) + \Theta({\cal
  E}^{(-+)}_0 - E_e)\,\frac{({\cal E}^{(-+)}_0 + m_e -
  E_e)\sqrt{({\cal E}^{(-+)}_0 - E_e)({\cal E}^{(-+)}_0 + 2m_e -
    E_e)}}{(E_0 - E_e)^2} \nonumber\\
\hspace{-0.3in}&&\times \Big( - \frac{2}{1 + 3\lambda^2}\,\big({\rm
  Re}(h_V\bar{h}^*_A) + |\bar{h}_A|^2\big) - A_0\,\frac{1}{1 +
  3\lambda^2}\,\big( |h_V|^2 + 3 |\bar{h}_A|^2\big)\Big)\Big\}
\,\frac{1}{2}\,\beta_eP_n (\cos\theta_1 + \cos\theta_2),
\end{eqnarray}
where $A^{(\rm SM)}(E_e)$ defines the electron asymmetry, calculated
in the SM (see Eqs.(17) and (20) of Ref.\cite{Ivanov2013}), $E_e =
E_-$ and $\beta_e = \beta_-$ are the energy and velocity of the decay
electron, $P_n = |\vec{\xi}_n| \le 1$ is the neutron spin
polarization. The electrons are detected in the solid angle $\Delta
\Omega^{(-)}_{12} = 2\pi (\cos\theta_1 - \cos\theta_2)$ with $0 \le
\varphi \le 2\pi$ and $\theta_1 \le \theta_- \le \theta_2$ with
respect to the neutron spin \cite{Dubbers2008} (see also
\cite{Abele2013,Abele2016}).  In case of non--relativistic
electron--positron pairs positrons can be detected in the solid angle
$\Delta \Omega^{(-)}_{12} = 2\pi (\cos\theta_1 - \cos\theta_2)$ in the
direction opposite to the direction of the electron 3--momentum
$\vec{k}_-$.

\section{Probability distribution of dark matter decay 
$n \to \chi + e^- + e^+$ as a function of invariant mass of
  electron--positron pairs}
\label{sec:mass}

\begin{figure}
\includegraphics[height=0.15\textheight]{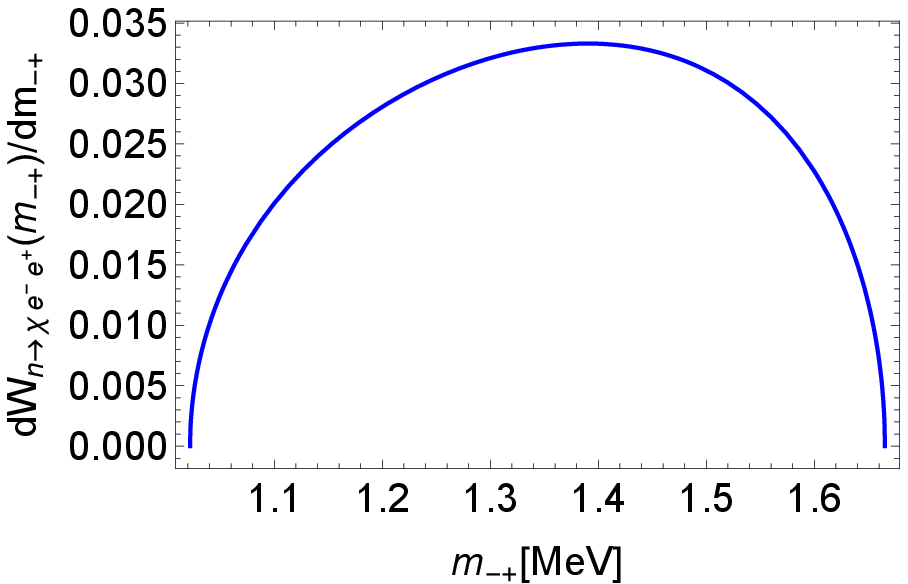}
\includegraphics[height=0.15\textheight]{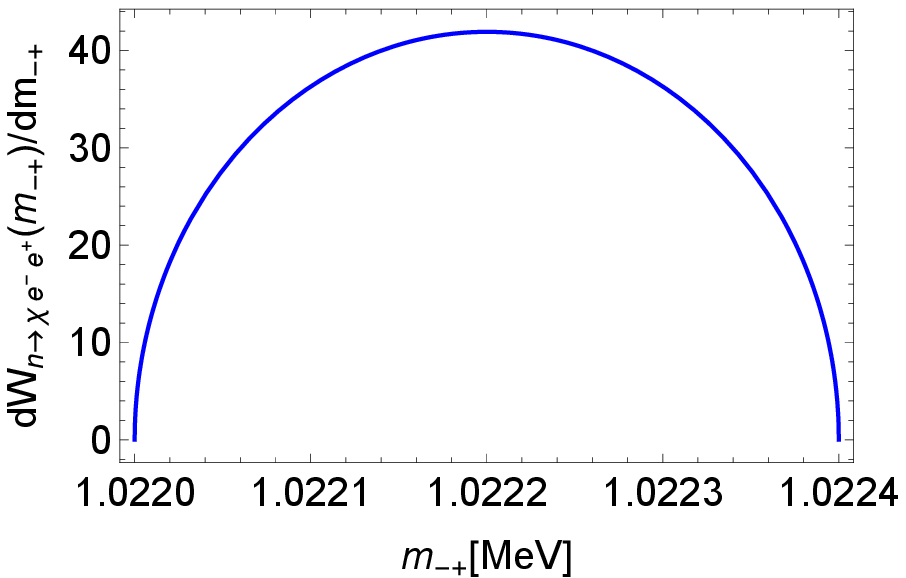}
  \caption{The probability distribution of the neutron dark matter
    decay $n \to \chi + e^- + e^+$ Eq.(\ref{eq:16}) as a function of
    the invariant mass of the electron--positron pair, where left and
    right figures are plotted for $a^{(\rm dm)} = 0$ and the mass
    $m_{\chi}$ of the dark matter fermion $m_{\chi} = 937.9\,{\rm
      MeV}$ and $m_{\chi} = 938.543\,{\rm MeV}$, respectively.}
\label{fig:fig3}
\end{figure}

The results, obtained in this section, can be also used for the
analysis of experimental data on searches of the dark matter decay
mode $n \to \chi + e^- + e^+$ in prospective experiments of the UCNA
and PERKEO Collaborations for electron--positron kinetic energies
$T_{-+} < 100\,{\rm keV}$.  The relativistic covariant calculation
gives for the density of the rate of the dark matter decay $n \to \chi
+ e^- + e^+$ the following expression
\begin{eqnarray}\label{eq:15}
\hspace{-0.3in}&&d^9\lambda_{n\to \chi\,e^-e^+} =
\frac{G^2_F|V_{ud}|^2}{32\pi^5 m_n}\,\Big\{\big(|h_V|^2 +
|\bar{h}_A|^2\big)\,\big[(k_{\chi}\cdot k_-)(k_n\cdot k_+) +
  (k_{\chi}\cdot k_+)(k_n\cdot k_-)\big] - m_n m_{\chi}\big(|h_V|^2 -
|\bar{h}_A|^2\big) (k_-\cdot k_+)\nonumber\\
\hspace{-0.3in}&&+ 2 m_n\,{\rm Re}(h_V
\bar{h}^*_A\big)\,\big[(k_{\chi}\cdot k_-)(\zeta_n\cdot k_+) +
  (k_{\chi}\cdot k_+)(\zeta_n\cdot k_-)\big] - 2\,{\rm Re}(h_V
\bar{h}^*_A\big)\,\big[(k_{\chi}\cdot k_-)(k_n\cdot k_+) -
  (k_{\chi}\cdot k_+)(k_n\cdot k_-)\big]\nonumber\\ \hspace{-0.3in}&&-
m_n \,\big(|h_V|^2 + |\bar{h}_A|^2\big)\,\big[(k_{\chi}\cdot
  k_-)(\zeta_n\cdot k_+) - (k_{\chi}\cdot k_+)(\zeta_n\cdot k_-)\big]
+ m_{\chi}\,\big(|h_V|^2 - |\bar{h}_A|^2\big)\,\big[(k_n \cdot
  k_-)(\zeta_n\cdot k_+)\nonumber\\ \hspace{-0.3in}&& - (k_n \cdot
  k_+)(\zeta_n\cdot k_-)\big] - m_{\chi} \,i \,\big(h_V \bar{h}^*_A -
h^*_V \bar{h}_A\big)\,\varepsilon^{\alpha\beta\mu\nu}k_{n\alpha}
\zeta_{n\beta} k_{-\mu}k_{+\nu} \Big\}\, \delta^{(4)}(k_n - k_{\chi} -
k_- - k_+)\,\frac{d^3k_{\chi}}{E_{\chi}}\,\frac{d^3k_-}{ E_-}\,
\frac{d^3k_+}{E_+}.
\end{eqnarray}
To leading order in the large heavy neutron and dark matter fermion
expansion the right--hand--side (r.h.s.) of Eq.(\ref{eq:15})
reproduces the electron--energy and angular distribution given by
Eq.(\ref{eq:6}).  Then, using Eq.(\ref{eq:15}) and following the
technique of the calculation of the phase--volume of three--particle
decays \cite{Byckling1975} we define the probability distribution of
the decay $n \to \chi + e^- + e^+$ with respect to the neutron
$\beta^-$--decay as a function of the invariant mass of the
electron--positron pair $m^2_{-+} = (k_- + k_+)^2$. We get
\begin{eqnarray}\label{eq:16}
\hspace{-0.3in}\frac{dW_{n\to \chi\,e^-e^+}(m_{-+})}{dm_{-+}} &=&
\frac{1}{48 m^3_n}\,\frac{9.68 \times 10^{-3}\sqrt{\big(m^2_{-+} -
    4m^2_e\big)\,\big((m^2_n - m^2_{\chi} + m^2_{-+})^2 - 4 m^2_n
    m^2_{-+}\big)}}{\displaystyle \int ^{{\cal
      E}^{(-+)}_0}_{m_e}({\cal E}^{(-+)}_0 + m_e - E_-)\sqrt{({\cal
      E}^{(-+)}_0 - E_-)({\cal E}^{(-+)}_0 + 2m_e - E_-)}\,\sqrt{E^2_-
    - m^2_e}\, E_-\,dE_-}\nonumber\\
\hspace{-0.3in}&&\times\, \Big\{\Big[(m^2_{-+} - 4 m^2_e)\,(m^2_n+
  m^2_{\chi} - m^2_{-+}) + \Big(1 + \frac{2
    m^2_e}{~~~m^2_{-+}}\Big)\,\big((m^2_n- m^2_{\chi})^2 -
  m^4_{-+}\big)\Big]\nonumber\\
\hspace{-0.3in}&& + a^{(\rm dm)} \Big[(m^2_{-+} - 4 m^2_e)\,(m^2_n+
  m^2_{\chi} - m^2_{-+}) - 12 m_n m_{\chi} \,(m^2_{-+} - 2 m^2_e) +
  \Big(1 + \frac{2 m^2_e}{~~~m^2_{-+}}\Big)\nonumber\\
\hspace{-0.3in}&&\times\,\big((m^2_n-
  m^2_{\chi})^2 - m^4_{-+}\big)\Big]\Big\},
\end{eqnarray}
where $W_{n \to \chi\,e^-\,e^+} = \tau_n\lambda_{n\to \chi\,e^-\,e^+}$
and ${\cal E}^{(-+)}_0 = (m^2_n - m^2_{\chi} - 2 m_{\chi}m_e)/2
m_n$. To leading order in the large neutron and dark matter fermion
mass expansion ${\cal E}^{(-+)}_0 = m_n - m_{\chi} - m_e$. We remind
that all masses in the r.h.s. of Eq.(\ref{eq:16}) are measured in
MeV. The invariant mass $m_{-+}$ varies in the limits $2m_e \le m_{-+}
\le (m_n - m_{\chi})$ \cite{Byckling1975}. Since to leading order in
the large neutron and dark matter fermion heavy mass expansion the
energy of the electron--positron pair $E_{-+}$ is equal to the mass
difference $E_{-+} = E_- + E_+ = m_n - m_{\chi}$, the squared
invariant mass $m^2_{-+}$ as a function of the energy difference $(E_-
- E_+)$ is defined by
\begin{eqnarray}\label{eq:17}
m^2_{-+} &=& (m_n - m_{\chi})^2 - \frac{1}{2}\Big((E_- - E_+)^2 + (m_n
- m_{\chi})^2 - 4 m^2_e\Big)\nonumber\\ &-& \frac{1}{2}
\sqrt{\big[(E_- - E_+)^2 + \big((m_n - m_{\chi})^2 -
    4m^2_e\big)\big]^2 - 4 (m_n - m_{\chi})^2\,(E_- -
  E_+)^2}\,\cos\vartheta.
\end{eqnarray}
where $\vec{k}_-\cdot \vec{k}_+ = k_-k_+\,\cos\vartheta$. Since for
$\vartheta = \pi$ and $\vartheta = 0$ we have to get $m^2_{-+} = (m_n
- m_{\chi})^2$ and $m^2_{+} = 4 m^2_e$ \cite{Byckling1975},
respectively, we derive that $E_- = E_+$. This gives
\begin{eqnarray}\label{eq:18}
m^2_{-+} = (m_n - m_{\chi})^2 \Big(1 - \Big(1 - \frac{4m^2_e}{(m_n -
  m_{\chi})^2}\Big)\,\cos^2\frac{\vartheta}{2}\Big).
\end{eqnarray}
In Fig.\,\ref{fig:fig3} we plot the probability density
Eq.(\ref{eq:16}) at $a^{(\rm dm)} = 0$ and for $m_{\chi} = 937.9\,{\rm
  MeV}$ (left figure) and $m_{\chi} = 938.543\,{\rm MeV}$ (right
figure) in the invariant mass region $2m_e \le m_{-+} \le (m_n -
m_{\chi})$, respectively. These plots may serve as
  theoretical backgrounds for the analysis of the experimental data on
  searches of the dark matter decay mode $n \to \chi + e^- + e^+$ in
  the experiments of the UCNA and PERKEO Collaborations at
  electron--positron kinetic energies $T_{-+} < 100\,{\rm keV}$.

\section{Low--energy electron--neutron inelastic scattering}
\label{sec:crosssection}

Because of the magnetic moment a neutron, moving with a 3-momentum
$\vec{k}_n$, can couple to an electric field of electrically charged
particles \cite{Schwinger1948,Dalitz1951,Gerasimov1963}. In turn, a
moving electrically charged particle such as an electron may also
couple to a neutron at rest. Below we calculate the differential cross
section for the reaction $e^- + n \to \chi + e^-$, which can be
measured above a background of a low--energy electron--neutron
electromagnetic interaction. The amplitude of the reaction $e^- + n
\to \chi + e^-$ to leading order in the large neutron and dark matter
fermion heavy mass expansion is equal to
\begin{eqnarray}\label{eq:19}
M(e^- n\to \chi e^-) = \sqrt{4 m_n m_{\chi}}\,\frac{G_F
  V_{ud}}{\sqrt{2}}\,\Big(h_V
[\varphi^{\dagger}_{\chi}\varphi_n][\bar{u}'_e \gamma^0(1 - \gamma^5)
  u_e] - \bar{h}_A [\varphi^{\dagger}_{\chi}\vec{\sigma}
  \varphi_n]\cdot [\bar{u}'_e \vec{\gamma}\,(1 - \gamma^5) u_e]\Big).
\end{eqnarray}
The hermitian conjugate amplitude is
\begin{eqnarray}\label{eq:20}
M^{\dagger}(e^- n\to \chi e^-) = \sqrt{4 m_n m_{\chi}}\,\frac{G_F
  V_{ud}}{\sqrt{2}}\,\Big(h^*_V
[\varphi^{\dagger}_n\varphi_n][\bar{u}_e \gamma^0(1 - \gamma^5) u'_e] -
\bar{h}^*_A [\varphi^{\dagger}_n\vec{\sigma} \varphi_{\chi}]\cdot [\bar{u}_e
  \vec{\gamma}\,(1 - \gamma^5) u'_e]\Big).
\end{eqnarray}
In Eqs.(\ref{eq:19}) and (\ref{eq:20}) $\varphi_{\chi}$ and
$\varphi_n$ are the Pauli wave functions of the dark matter fermion
and neutron, respectively, and $u'_e$ and $u_e$ are the Dirac wave
functions of free electrons in the final and initial states of the
reaction $e^- + n \to \chi + e^-$. The differential cross section for
the reaction $e^- + n \to \chi + e^-$ with a polarized neutron is
equal to
\begin{eqnarray}\label{eq:21}
\frac{d^2\sigma(\vec{\xi}_n, \vec{k}^{\,'}_e, \vec{k}_e)}{d\Omega'_e}
&=& (1 + 3 \lambda^2)\,\frac{G^2_F|V_{ud}|^2}{8\pi^2}\,E'_e
E_e\,\frac{k'_e}{k_e}\,\zeta^{(\rm dm)}\,\Big(1 + a^{(\rm
  dm)}\,\frac{\vec{k}^{\,'}_e\cdot \vec{k}_e }{E'_e E_e} + A^{(\rm
  dm)}_+\,\frac{\vec{\xi}_n \cdot \vec{k}^{\,'}_e}{E'_e} + A^{(\rm
  dm)}_-\,\frac{\vec{\xi}_n \cdot \vec{k}_e}{E_e}\Big) =
\nonumber\\ &=& \frac{\pi}{4}\, \frac{\zeta^{(\rm dm)}}{\tau_n
  f_n}\,E'_e E_e\,\frac{k'_e}{k_e}\,\Big(1 + a^{(\rm
  dm)}\,\frac{\vec{k}^{\,'}_e\cdot \vec{k}_e }{E'_e E_e} + A^{(\rm
  dm)}_+\,\frac{\vec{\xi}_n \cdot \vec{k}^{\,'}_e}{E'_e} + A^{(\rm
  dm)}_-\,\frac{\vec{\xi}_n \cdot \vec{k}_e}{E_e}\Big),
\end{eqnarray}
where $\vec{\xi}_n$ is a unit vector of the neutron polarization,
$E'_e = {\cal E}_0 + E_e$ with ${\cal E}_0 = m_n - m_{\chi}$ and
$d\Omega'_e = \sin\theta'_e d\theta'_e d\phi'_e$ is the infinitesimal
solid angle of the 3--momentum $\vec{k}^{\,'}_e$ of the outgoing
electron such as $\vec{k}^{\,'}_e\cdot \vec{k}_e = k'_e
k_e\big(\cos\theta'_e \cos\theta_e + \sin\theta'_e \sin\theta_e
\cos(\phi'_e - \phi_e)\big)$. The correlation coefficients are defined
in Eq.(\ref{eq:7}). The cross section for the reaction $e^- + n \to
\chi + e^-$ with a polarized neutron is
\begin{eqnarray}\label{eq:22}
\sigma(\vec{\xi}_n, \vec{k}_e) = \frac{\pi}{4}\, \frac{\zeta^{(\rm
    dm)}}{\tau_n f_n}\,E'_e E_e\,\frac{k'_e}{k_e}\,\Big(1 + A^{(\rm
  dm)}_-\,\frac{\vec{\xi}_n \cdot \vec{k}_e}{E_e}\Big).
\end{eqnarray}
The differential cross section for the reaction $e^- + n \to \chi +
e^-$ with an unpolarized neutron is given by
\begin{eqnarray}\label{eq:23}
\frac{d^2\sigma(k_e, \vec{n}\,',\vec{n}\,)}{d\Omega'_e} =
\frac{\pi}{4}\, \frac{\zeta^{(\rm dm)}}{\tau_n f_n}\,E'_e
E_e\,\frac{k'_e}{k_e}\,\Big(1 + a^{(\rm dm)}\,\frac{k'_ek_e }{E'_e
  E_e}\,\vec{n}\,' \cdot \vec{n}\,\Big),
\end{eqnarray}
where $\vec{n}\,'$ and $\vec{n}$ are unit vectors in the directions
$\vec{k}^{\,'}_e$ and $\vec{k}_e$, respectively. For low--energy
electrons the differential cross section Eq.(\ref{eq:23}) can be
transcribed into the form
\begin{eqnarray}\label{eq:24}
\frac{d^2\sigma(\beta_e,\vec{n}\,',\vec{n}\,)}{d\Omega'_e} =
\frac{\sigma_0}{4\pi \beta_e}\,\Big(1 + a^{(\rm dm)}\,\frac{\sqrt{(m_n
    - m_{\chi})(m_n - m_{\chi} + 2m_e)}}{m_n - m_{\chi} +
  m_e}\,\beta_e\,\vec{n}\,' \cdot \vec{n}\,\Big),
\end{eqnarray}
where we have used the definition of the coupling constant
$\zeta^{(\rm dm)}$ in Eq.(\ref{eq:13}), $k'_e = \sqrt{{\cal E}_0(
  {\cal E}_0 + 2m_e)}$ and $\beta_e$ is the velocity of the incoming
electron. Then, $\sigma_0$ is equal to
\begin{eqnarray}\label{eq:25}
\sigma_0 = 0.84\times 10^{-21}\,\frac{(m_n - m_{\chi} + m_e)\sqrt{m_n -
    m_{\chi} + 2 m_e}}{(m_n - m_{\chi})^{9/2}}\,{\rm b},
\end{eqnarray}
where the fermion masses are measured in MeV.  If $m_n - m_{\chi} \le
0.01\,{\rm MeV}$ we get $\sigma_0 \ge 0.45\,{\rm pb}$. For the kinetic
energy of incoming electrons $T_e \simeq 1\,{\rm eV}$ we get $\beta_e
\simeq 2\times 10^{-3}$ and
\begin{eqnarray}\label{eq:26}
4\pi
\frac{d^2\sigma(\beta_e,\vec{n}\,',\vec{n}\,)}{d\Omega'_e}\Big|_{\beta_e
  \simeq 2\times 10^{-3}} \simeq 0.22\,\Big(1 + 4\times
10^{-4}\,a^{(\rm dm)}\,\vec{n}\,' \cdot \vec{n}\,\Big) {\rm nb}.
\end{eqnarray}
This differential cross section possesses the following properties: i)
it is practically isotropic and ii) for $m_n - m_{\chi} \le 0.01\,{\rm
  MeV}$ the momenta of outgoing electrons are restricted by the values
$k'_e \le 0.10\,{\rm MeV}$ which are much larger than the momenta $k_e
\simeq 10^{-3}\,{\rm MeV}$ of incoming electrons. In turn, using $m_n
- m_{\chi} \ge 0.12\,{\rm MeV}$ (see discussion below
Eq.(\ref{eq:51})) for the r.h.s. of Eq.(\ref{eq:26}) we get $ 3.95\,(1
+ 10^{-3}\,a^{(\rm dm)}\,\vec{n}\,' \cdot \vec{n}\,)\,{\rm fb}$ and
$k'_e \ge 0.37\,{\rm MeV}$, respectively.  In spite of a sufficiently
small value the differential cross section for the inelastic
electron--neutron scattering Eq.(\ref{eq:26}) can be distinguished
above the background defined by the differential cross section for the
reaction $e^- + n \to n + e^-$ caused by the electromagnetic
electron--neutron coupling \cite{Hofstadter1956,Yennie1957}.

\section{Gauge invariant quantum field theory model for UV completion
 of effective interaction Eq.(\ref{eq:5}) and dark matter dynamics in
 neutron stars}
\label{sec:uvcompletion}

In this section we propose a gauge invariant quantum field theory
model for the UV completion of the effective interaction
Eq.(\ref{eq:5}) and dark matter dynamics in neutron stars. We discuss
compatibility of the predictions of this model with constraints on i)
the dark matter production in ATLAS experiments at the LHC, ii) the
cross section for low--energy dark matter fermion--electron scattering
and iv) dark matter properties following from interference of dark
matter into dynamics of neutron stars.

We construct a gauge invariant quantum field theory model of nucleon,
electron and neutrino and dark matter particles with $SU_L(2)\times
U_R(1)\times U'_R(1) \times U''_L(1)$ gauge symmetry. Such a quantum
field theory model contains the sector of the Standard Electroweak
Model (SEM) (or the SM sector) \cite{PDG2018} with $SU_L(2)\times
U_R(1)$ gauge symmetry and the dark matter sector with $U'_R(1) \times
U''_L(1)$ gauge symmetry. The dark matter sectors invariant under
$U'_R(1)$ and $U''_L(1)$ gauge symmetries are responsible for the UV
completion of the effective interaction Eq.(\ref{eq:5}) and
interference of the dark matter into dynamics of neutron stars,
respectively. First, we construct the quantum field theory model
invariant under gauge $SU_L(2)\times U_R(1)\times U'_R(1)$
transformations for the UV completion of the effective interaction
Eq.(\ref{eq:5}), and then we extend it by the dark matter sector
invariant under gauge $U''_L(1)$ symmetry, responsible for dark matter
dynamics in neutron stars.

\subsection{Quantum field theory model with gauge $SU_L(2) \times U_R(1) 
\times U'_R(1)$ symmetry for UV--completion of effective interaction
Eq.(\ref{eq:5})}

The SM sector of our model, including left-- and right--handed
neutron, proton, and electron and left--handed neutrino, is invariant
under gauge $SU_L(2)\times U_R(1)$ transformations and possesses
symmetric and spontaneously broken (or physical) phases. In the
symmetric phase the interactions are mediated by two gauge fields
$\vec{W}_{\mu} = (W^1_{\mu},W^2_{\mu},W^3_{\mu})$ and $B_{\nu}$ and a
doublet of the Higgs--field $\phi$. The covariant derivatives of the
matter fields and the Higgs--field are defined by \cite{PDG2018}
\begin{eqnarray}\label{eq:27}
D_{L\mu} =\partial_{\mu} + i\,g'\,\frac{1}{2}\,Y\, B_{\mu}+ i\,g\,
\vec{t}\cdot \vec{W}_{\mu}\quad,\quad  D_{R\mu} =
\partial_{\mu} + i\,g'\,\frac{1}{2}\,Y\, B_{\mu},
\end{eqnarray}
where matrix $\vec{t}\,$ is the $2\times 2$ matrix of weak isospin,
defined in terms of the Pauli matrices $\vec{\tau}$ as $\vec{t} =
\frac{1}{2}\,\vec{\tau}$, such as $[t^a,t^b] = i\varepsilon^{abc}t^c$,
and $Y$ is the operator of the weak hypercharge, related to the
operators of the weak isospin $\vec{I} = \vec{t}$ and electric charge
as $Q = I_3 + \frac{Y}{2}$ \cite{PDG2018}. Then, $g$ and $g'$ are
gauge coupling constants. The Lagrangian of the SM sector in the
symmetric phase takes the form
\begin{eqnarray}\label{eq:28}
\hspace{-0.3in}&&{\cal L}_{\rm SM} = -
\frac{1}{4}\,\vec{W}_{\mu\nu}\cdot \vec{W}^{\mu\nu} -
\frac{1}{4}\,B_{\mu\nu} B^{\mu\nu} +
\bar{\Psi}_{NL}i\gamma^{\mu}D_{L\mu}\Psi_{NL} +
\bar{\Psi}_{eL}i\gamma^{\mu}D_{L\mu}\Psi_{eL}\nonumber\\\hspace{-0.3in}
&& + \bar{\psi}_{pR}i\gamma^{\mu}D_{R\mu}\psi_{pR} +
\bar{\psi}_{nR}i\gamma^{\mu}D_{R\mu}\psi_{nR} +
\bar{\psi}_{eR}i\gamma^{\mu}D_{R\mu}\psi_{eR} +
(D_{L\mu}\phi)^{\dagger}D^{\mu}_{L}\phi + \tilde{\mu}^2\,
\phi^{\dagger}\phi -
\tilde{\lambda}\,(\phi^{\dagger}\phi)^2\nonumber\\
\hspace{-0.3in}&& - g_p(\bar{\Psi}_{NL}\psi_{pR}\phi^c + \phi^{c
  \dagger}\bar{\psi}_{pR}\Psi_{NL}) - g_n(\bar{\Psi}_{NL}\psi_{nR}\phi
+ \phi^{\dagger}\bar{\psi}_{nR}\Psi_{NL}) -
g_e(\bar{\Psi}_{eL}\psi_{eR}\phi +
\phi^{\dagger}\bar{\psi}_{eR}\Psi_{eL}).
\end{eqnarray}
Here the field strength tensor operators of the gauge fields
$\vec{W}_{\mu\nu}$ and $B_{\mu\nu}$ are defined by
\begin{eqnarray}\label{eq:29}
\vec{W}_{\mu\nu} = \partial_{\mu}\vec{W}_{\nu} -
\partial_{\nu}\vec{W}_{\mu} - g \vec{W}_{\mu}\times
\vec{W}_{\nu}\quad, \quad B_{\mu\nu} = \partial_{\mu}B_{\nu} -
\partial_{\nu}B_{\mu},
\end{eqnarray}
$\Psi_{NL}$, $\Psi_{eL}$ and $\phi$ are operators of the left--handed
nucleon and lepton field doublets and the Higgs--field, respectively,
\begin{eqnarray}\label{eq:30}
\Psi_{NL} = P_L\left(\begin{array}{c}\psi_p \\ \psi_n
\end{array}\right) \quad,\quad \Psi_{eL}  =  P_L\left(\begin{array}{c} 
\psi_{\nu_e} \\ \psi_e
\end{array}\right)\quad,\quad \phi  = \left(\begin{array}{c}\phi^+
    \\ \phi^0
\end{array}\right)
\end{eqnarray}
and $\psi_{pR}$, $\psi_{nR}$ and $\psi_{eR}$ are operators of the
proton, neutron and electron right--handed fields
\begin{eqnarray}\label{eq:31}
\psi_{pR} = P_R\psi_p \quad,\quad \psi_{nR} = P_R\psi_n \quad,\quad
\psi_{eR} = P_R\psi_e,
\end{eqnarray}
where $P_{L,R} = (1 \mp\gamma^5)/2$ are the projection operators;
$P^2_{L,R}= P_{L,R}$ and $P_LP_R = P_RP_L = 0$, and $\phi^c =
i\tau^2\phi^*$.  The coupling constants $g_p$, $g_n$ and $g_e$ are
connected with the masses of the proton, neutron and electron. The
parameters $\tilde{\mu}^2$ and $\tilde{\lambda}$ define a
non--vanishing vacuum expectation value of the Higgs--field $\phi$
corresponding to a physical phase of the system described by the
Lagrangian Eq.(\ref{eq:27}).

In the physical phase the components of the Higgs--field $\phi$ are
equal to $\phi^+ = 0$ and $\phi^0 = (v + \varphi)/\sqrt{2}$, 
respectively, where $v$ is the vacuum expectation value $\langle
\phi^0\rangle$ and $\varphi$ is an observable scalar Higgs--field. In
the physical phase the Lagrangian Eq.(\ref{eq:27}) describes massive
electroweak boson $(W^{\mp}_{\mu},Z_{\mu})$ and massless
electromagnetic $A_{\mu}$ fields, where $Z_{\mu}$ and $A_{\mu}$ are
linear superpositions of the $W^3_{\mu}$ and $B_{\mu}$ gauge fields, a
massive scalar Higgs--field $\varphi$, massive proton, neutron and
electron fields, and a massless neutrino field, respectively.

Now for the UV completion of the effective interaction Eq.(\ref{eq:5})
we have to introduce the dark matter sector. For this aim together
with the right--handed dark matter fermion field $\chi$, described by
the field operator $\psi_{\chi R}$, we introduce the dark matter
spin--1 $C_{\mu}$ and complex scalar boson $\Phi$ fields. The
Lagrangian of the dark matter sector invariant under gauge $U'_R(1)$
transformations takes the following form
\begin{eqnarray}\label{eq:32}
&&{\cal L}_{\rm DM'} = \bar{\psi}_{\chi R}i\gamma^{\mu}(\partial_{\mu}
  + i e_{\chi}C_{\mu})\psi_{\chi R} -
  \frac{1}{4}\,C_{\mu\nu}C^{\mu\nu} + (\partial_{\mu} -
  ie_{\chi}C_{\mu})\Phi^*(\partial_{\mu} + ie_{\chi}C_{\mu})\Phi +
  \kappa^2|\Phi|^2 - \gamma |\Phi|^4 \nonumber\\ &&+ \bar{\psi}_{\chi
    L}i\gamma^{\mu}\partial_{\mu}\psi_{\chi L} - \sqrt{2}\,
  f_{\chi}\big(\bar{\psi}_{\chi R}\psi_{\chi L} \Phi +
  \bar{\psi}_{\chi L}\psi_{\chi R} \Phi^*\big) +
  \bar{\Psi}_{eL}i\gamma^{\mu} \big(\ldots + ie_{\chi}
  C_{\mu}\big)\Psi_{eL} - 2 \zeta_e (\bar{\Psi}_{eL}\psi_{eR}\phi\,\Phi +
  \Phi^*\phi^{\dagger}\bar{\psi}_{eR}\Psi_{eL})\nonumber\\ &&+ 2
  \sqrt{2}\,\xi_{\chi}\big( \Phi^*\bar{\psi}_{n
    R}i\gamma^{\mu}(\partial_{\mu} + i e_{\chi}C_{\mu})\psi_{\chi R} -
  i(\partial_{\mu} - i e_{\chi}C_{\mu})\bar{\psi}_{\chi R}\gamma^{\mu}
  \psi_{n R} \Phi\big),
\end{eqnarray}
where $C_{\mu\nu} = \partial_{\mu}C_{\nu} - \partial_{\nu}C_{\mu}$ is
the field strength tensor operator of the dark matter spin--1 field
$C_{\mu}$, $e_{\chi}$ is a gauge coupling constant or the dark matter
``charge'' of the right--handed dark matter fermion $\chi$ and the
left--handed SM electron and neutrino. The parameters $\kappa^2$ and
$\gamma$ define a non--vanishing vacuum expectation value of the dark
matter scalar field $\Phi$, that leads to a non--vanishing mass
$m_{\chi}$ of the dark matter fermion field $\psi_{\chi}$, which
should be proportional to the coupling constant $f_{\chi}$. Then, the
coupling constant $\xi_{\chi}$ defines a mixing of the right--handed
neutron with the right--handed dark matter fermion. In the term
$\bar{\Psi}_{eL}i\gamma^{\mu} \big(\ldots + ie_{\chi}
C_{\mu}\big)\Psi_{eL}$ the ellipsis denotes the covariant derivative
$D_{L\mu}$. This means that the covariant derivative of the
left--handed leptons in the quantum field theory, described by the
Lagrangian ${\cal L}_{\rm SM + DM'} = {\cal L}_{\rm SM} + {\cal
  L}_{\rm DM'}$, should be taken in the form $D_{L\mu} + i e_{\chi}
C_{\mu}$. We have also redefined the electron mass term
$g_e(\bar{\Psi}_{eL}\psi_{eR}\phi +
\phi^{\dagger}\bar{\psi}_{eR}\Psi_{eL}) \to 2 \zeta_e
(\bar{\Psi}_{eL}\psi_{eR}\phi\,\Phi +
\Phi^*\phi^{\dagger}\bar{\psi}_{eR}\Psi_{eL})$. The Lagrangian
Eq.(\ref{eq:32}) is invariant under $U'_R(1)$ dark matter gauge
transformations
\begin{eqnarray}\label{eq:33}
\psi_{\chi R} \to \psi'_{\chi R} = e^{\,i\alpha_{\chi}}\psi_{\chi
  R}\quad,\quad \Phi \to \Phi' = e^{\,i\alpha_{\chi}}\Phi \quad,\quad
\Psi_{eL} \to \Psi'_{eL} = e^{\,i\alpha_{\chi}} \Psi_{eL}\quad,\quad
C_{\mu} \to C'_{\mu} = C_{\mu} -
\frac{1}{e_{\chi}}\,\partial_{\mu}\alpha_{\chi},
\end{eqnarray}
where $\alpha_{\chi}$ is a gauge parameter.  We would like to notice
that the right--handed neutron field, described by the field operator
$\psi_{n R}$, and the left--handed dark matter fermion field
$\psi_{\chi L}$ are invariant under $SU_L(2)\times U_R(1) \times
U'_R(1)$ gauge transformations.  In turn, the right--handed electron
field operator $\psi_{eR}$ is invariant under $U'_R(1)$ dark matter
gauge transformations. In order to define the dark matter sector in
the physical phase we take the complex dark matter scalar field $\Phi$
in the following form $\Phi = e^{i\,\alpha_{\chi}}\rho/\sqrt{2}$
\cite{Kibble1967,Kibble2015}, where $\rho$ is a dark matter scalar
field, and make a gauge transformation $\psi_{\chi R} \to
e^{i\,\alpha_{\chi}}\psi_{\chi R}$ and $\Psi_{eL} \to
e^{i\,\alpha_{\chi}}\Psi_{eL}$ \cite{Kibble1967,Kibble2015}. As a
result we arrive at the Lagrangian
\begin{eqnarray}\label{eq:34}
\hspace{-0.3in}{\cal L}_{\rm DM'} &=& \bar{\psi}_{\chi
  R}i\gamma^{\mu}(\partial_{\mu} + i \partial_{\mu}\alpha_{\chi} + i
e_{\chi}C_{\mu})\psi_{\chi R} + \bar{\psi}_{\chi
  L}i\gamma^{\mu}\partial_{\mu} \psi_{\chi L} -
f_{\chi}\bar{\psi}_{\chi}\psi_{\chi}\rho -
\frac{1}{4}\,C_{\mu\nu}C^{\mu\nu} + \frac{1}{2}\partial_{\mu}\rho
\partial^{\mu}\rho \nonumber\\ \hspace{-0.3in}&+& \frac{1}{2}(
\partial_{\mu}\alpha_{\chi} + e_{\chi}C_{\mu})(
\partial^{\mu}\alpha_{\chi} + e_{\chi}C^{\mu})\rho^2 +
\frac{1}{2}\,\kappa^2\rho^2 - \frac{1}{4}\,\gamma \rho^4 +
\bar{\Psi}_{eL}i\gamma^{\mu} \big(\ldots + i
\partial_{\mu}\alpha_{\chi} + ie_{\chi} C_{\mu}\big)\Psi_{eL}
\nonumber\\ \hspace{-0.3in}&+& 2\,\xi_{\chi}\rho\,\big(\bar{\psi}_{n
  R}i\gamma^{\mu}(\partial_{\mu} + i \partial_{\mu}\alpha_{\chi} + i
e_{\chi}C_{\mu})\psi_{\chi R} - i(\partial_{\mu} - i
\partial_{\mu}\alpha_{\chi} - i e_{\chi}C_{\mu})\bar{\psi}_{\chi
  R}\gamma^{\mu} \psi_{n R}\big)\nonumber\\ \hspace{-0.3in}&-&
\sqrt{2}\,\zeta_e\, \rho\, (\bar{\Psi}_{eL} \psi_{eR}\phi +
\phi^{\dagger}\bar{\psi}_{eR}\Psi_{eL}).
\end{eqnarray}
The field $C_{\mu} + \partial_{\mu}\alpha_{\chi}/e_{\chi}$ can be
treated as a new dark matter spin--1 $Z'$ field \cite{Kibble1967}. In
terms of the $Z'$--field the Lagrangian Eq.(\ref{eq:34}) reads
\begin{eqnarray}\label{eq:35}
{\cal L}'_{\rm DM} &=&
\bar{\psi}_{\chi}i\gamma^{\mu}\partial_{\mu}\psi_{\chi} -
e_{\chi}\bar{\psi}_{\chi R}\gamma^{\mu}\psi_{\chi R} Z'_{\mu} -
\frac{1}{4}\,Z'_{\mu\nu}Z'^{\mu\nu} + \frac{1}{2}\,e^2_{\chi}Z'_{\mu}
Z'^{\mu}\rho^2 - f_{\chi}\,\bar{\psi}_{\chi}\psi_{\chi} \rho +
\frac{1}{2}\partial_{\mu}\rho\partial^{\mu}\rho\nonumber\\ &+&
\frac{1}{2}\,\kappa^2 \rho^2 - \frac{1}{4}\,\gamma \rho^4 + 2
\,\xi_{\chi} \rho\,\big(\bar{\psi}_{n R}i\gamma^{\mu}(\partial_{\mu} +
i e_{\chi} Z'_{\mu}) \psi_{\chi R} - i(\partial_{\mu} - i e_{\chi}
Z'_{\mu})\bar{\psi}_{\chi R}\gamma^{\mu} \psi_{n
  R}\big)\nonumber\\ &+& \bar{\Psi}_{eL}i\gamma^{\mu} \big(\ldots +
ie_{\chi} Z'_{\mu}\big)\Psi_{eL} - \sqrt{2}\,\zeta_e \,\rho\,
(\bar{\Psi}_{eL} \psi_{eR}\phi +
\phi^{\dagger}\bar{\psi}_{eR}\Psi_{eL}).
\end{eqnarray}
where $Z'_{\mu\nu} = \partial_{\mu}Z'_{\nu} - \partial_{\nu}Z'_{\mu}$
is the field strength tensor operator of the dark matter spin--1 field
$Z'$.  The potential of the dark matter scalar $\rho$--field $V(\rho)
= - \frac{1}{2}\,\kappa^2 \rho^2 + \frac{1}{4}\,\gamma \rho^4$
possesses a minimum at $\langle \rho \rangle = v_{\chi} =
\sqrt{\kappa^2/\gamma}$. Introducing a new scalar field $\rho =
v_{\chi} + \sigma$ \cite{Kibble1967} we transcribe the Lagrangian
${\cal L}_{\rm SM + DM'} = {\cal L}_{\rm SM } + {\cal L}_{\rm DM'}$
into the form
\begin{eqnarray}\label{eq:36}
&&{\cal L}_{\rm SM + DM'} =
  \bar{\psi}_{\chi}\big(i\gamma^{\mu}\partial_{\mu} -
  m_{\chi}\big)\psi_{\chi} + \bar{\psi}_n
  \big(i\gamma^{\mu}\partial_{\mu} - m_n\big) \psi_n -
  e_{\chi}\bar{\psi}_{\chi R}\gamma^{\mu}\psi_{\chi R} Z'_{\mu} -
  \frac{1}{4}\,Z'_{\mu\nu} Z'^{\mu\nu} + \frac{1}{2}\,M^2_{Z'}Z'_{\mu}
  Z'^{\mu}\nonumber\\ &&+
  g_{\chi}\big(\bar{\psi}_ni\gamma^{\mu}(\partial_{\mu} + i e_{\chi}
  Z'_{\mu})(1 + \gamma^5)\psi_{\chi} - i(\partial_{\mu} - i e_{\chi}
  Z'_{\mu})\bar{\psi}_{\chi}\gamma^{\mu}(1 + \gamma^5) \psi_n\big) -
  e_{\chi} \bar{\Psi}_{eL}\gamma^{\mu}\Psi_{eL} Z'_{\mu}
  \nonumber\\ &&- \zeta_e \,(v_{\chi} + \sigma)\, (\bar{\Psi}_{eL}
  \psi_{eR}\phi + \phi^{\dagger}\bar{\psi}_{eR}\Psi_{eL}) +
  \xi_{\chi} e_{\chi}\,\sigma\,\big(\bar{\psi}_ni\gamma^{\mu}(\partial_{\mu}
  + i e_{\chi} Z'_{\mu})(1 + \gamma^5)\psi_{\chi} - i(\partial_{\mu} -
  i e_{\chi} Z'_{\mu})\bar{\psi}_{\chi}\gamma^{\mu}(1 + \gamma^5)
  \psi_n\big)\nonumber\\ &&+ e^2_{\chi} v_{\chi} Z'_{\mu}
  Z'^{\mu}\,\sigma + \frac{1}{2}\,e^2_{\chi} Z'_{\mu}
  Z'^{\mu}\,\sigma^2 -
  \sqrt{2}\,f_{\chi}\bar{\psi}_{\chi}\psi_{\chi}\sigma +
  \frac{1}{2}\partial_{\mu}\sigma\partial^{\mu}\sigma -
  \frac{1}{2}\,m^2_{\sigma}\,\sigma^2 - \gamma v_{\chi}\sigma^3 -
  \frac{1}{4}\,\gamma \sigma^4 + \ldots,
\end{eqnarray}
where $g_{\chi} = \xi_{\chi} v_{\chi}$ and $m_{\chi}$, $M_{Z'}$ and
$m_{\sigma}$ are masses of the dark matter fermion $\chi$, dark matter
spin--1 $Z'$ and dark matter scalar $\sigma$ fields
\begin{eqnarray}\label{eq:37}
m_{\chi} = f_{\chi}\,v_{\chi}\quad,\quad M_{Z'} = e_{\chi} v_{\chi}
\quad,\quad m_{\sigma} = \sqrt{2 \gamma} v_{\chi}.
\end{eqnarray}
In Eq.(\ref{eq:36}) the ellipsis denotes the contributions of other
kinetic and interaction terms, which can be obtained in the physical
phases of the SM sector described by the Lagrangian
Eq.(\ref{eq:28}). In principle, a mass of the dark matter scalar
$\sigma$--field is arbitrary. In order to allow the $n \to \chi$
transitions only by virtue of the dark matter spin--1 boson $Z'$ we
may delete the $\sigma$--field from its interactions taking the limit
$m_{\sigma} \to \infty$. This agrees well with the
Appelquist--Carazzone decoupling theorem
\cite{Appelquist1975}. Indeed, keeping the ratio $v_{\chi} =
\sqrt{\kappa^2/\gamma}$ fixed one may set $\gamma \to \infty$. This is
similar to the removal of the scalar $\sigma$--meson from its
interactions in the linear $\sigma$--model of strong low--energy
interactions \cite{GellMann1960}--\cite{Ivanov2018b}.  The effective
interaction Eq.(\ref{eq:5}) can be reproduced by the following part of
the Lagrangian Eq.(\ref{eq:36})
\begin{eqnarray}\label{eq:38}
{\cal L}_{\rm n\chi \ell} &=& g_{\chi}\big(\bar{\psi}_ni\gamma^{\mu}(1
+ \gamma^5)\partial_{\mu}\psi_{\chi} -
\partial_{\mu}\bar{\psi}_{\chi}i\gamma^{\mu}(1 + \gamma^5)\psi_n\big)
- g_{\chi}e_{\chi}\big(\bar{\psi}_n\gamma^{\mu}(1 +
\gamma^5)\psi_{\chi} + \bar{\psi}_{\chi}\gamma^{\mu}(1 +
\gamma^5)\psi_n\big)Z'_{\mu}\nonumber\\ &-&
\frac{1}{2}\,e_{\chi}\bar{\psi}_{\chi}\gamma^{\mu}(1 + \gamma^5)
\psi_{\chi} Z'_{\mu} - \frac{1}{2}\,e_{\chi} \bar{\Psi}_e\gamma^{\mu}(1 -
\gamma^5) \Psi_e Z'_{\mu}.
\end{eqnarray}
This Lagrangian we use also for the analysis of compatibility of
predictions of our model with constraints on i) the dark matter
production in experiments of the ATLAS Collaboration and ii) the cross
sections for low--energy dark matter fermion--electron scattering.

\subsection{Self--energy corrections to the neutron state}

\begin{figure}
\centering \includegraphics[height=0.018\textheight]{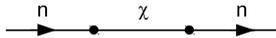}
  \caption{The Feynman diagrams, describing the contribution of the $n
    \leftrightarrow \chi$ transitions to the neutron mass.}
\label{fig:fig4}
\end{figure}

The first two terms of Eq.(\ref{eq:38}) define the self--energy
corrections to the neutron state. The contributions of the second term
are divergent are divergent and can be removed by renormalization of
the mass and wave function of the neutron. In turn, the contributions
of the first term are finite and caused by direct transitions of
right--handed neutron (dark matter fermion) to the right--handed dark
matter fermion (neutron). It is defined by the Feynman diagram in
Fig.\,\ref{fig:fig4}.
Having removed the contributions of the second term by renormalization
of the mass and wave function of the neutron we propose to estimate
the coupling constant $g_{\chi}$ from the contribution of the first
term in Eq.(\ref{eq:38}) to the neutron mass. Skipping intermediate
calculations we get
\begin{eqnarray}\label{eq:39}
\delta m_{n \leftrightarrow \chi} = \frac{ 2 g^2_{\chi} m^3_n}{m^2_n -
  m^2_{\chi}} \simeq \frac{g^2_{\chi}m^2_n}{m_n - m_{\chi}},
\end{eqnarray}
where we have taken into account that $m_{\chi} \lesssim
m_n$. According to \cite{PDG2018}, the neutron mass is equal to $m_n =
939.565413(6)\,{\rm MeV}$. This means that the mass correction $\delta
m_{n \leftrightarrow \chi}$ should be smaller than $\delta m_{n
  \leftrightarrow \chi} < 6\times 10^{-6}\,{\rm MeV}$. This gives the
following constraint on the coupling constant $g_{\chi}$: $|g_{\chi}|
< 2.45\times 10^{-3}\,\sqrt{m_n - m_{\chi}}/m_n$, where masses are
measured in MeV. The coupling constant $g_{\chi}$ is
dimensionless. According to Babu and Mohapatra \cite{Babu2015} the
first two terms in the Lagrangian Eq.(\ref{eq:38}) should induce also
$n \leftrightarrow \chi$ oscillations (see Eq.(14) of
Ref.\cite{Babu2015}). This effect demands a special analysis, which
goes beyond the scope of this paper.

\subsection{UV completion for effective interaction Eq.(\ref{eq:5})}

\begin{figure}
\centering \includegraphics[height=0.075\textheight]{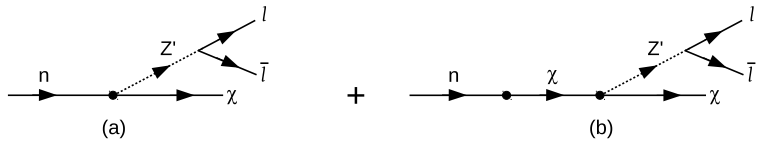}
  \caption{The Feynman diagrams, describing the amplitude of the
    neutron dark matter decay $n \to \chi + \ell + \bar{\ell}$.}
\label{fig:fig5}
\end{figure}

The amplitude of the dark matter decays $n \to \chi + \ell +
\bar{\ell}$, where $\ell = e^-, \nu_e$ and $\bar{\ell} = e^+,
\bar{\nu}_e$, respectively, is defined by the Feynman diagrams in
Fig.\,\ref{fig:fig5}.
The analytical expressions for the amplitudes of the decays $n \to
\chi + \ell + \bar{\ell}$ are given by
\begin{eqnarray}\label{eq:40}
M(n \to \chi + \ell + \bar{\ell})_{\rm Fig.\,\ref{fig:fig5}a} &=&
\frac{g_{\chi}e^2_{\chi}}{2}[\bar{u}_{\chi}(\vec{k}_{\chi},
  \sigma_{\chi})\gamma^{\mu}(1 + \gamma^5) u_n(\vec{k}_n,
  \sigma_n)]\,\frac{\displaystyle - \eta_{\mu\nu} +
  \frac{q_{\mu}q_{\nu}}{M^2_{Z'}}}{M^2_{Z'} - q^2 -
  i0}\nonumber\\ &&\times\,[\bar{u}_{\ell}(\vec{k}_{\ell})
  \gamma^{\nu}(1 - \gamma^5)v_{\bar{\ell}}(\vec{k}_{\bar{\ell}},
  \sigma_{\bar{\ell}})]
\end{eqnarray}
and 
\begin{eqnarray}\label{eq:41}
M(n \to \chi + \ell + \bar{\ell})_{\rm Fig.\,\ref{fig:fig5}b} &=&
- \frac{g_{\chi}e^2_{\chi}}{2}\,\frac{m^2_n}{m^2_n -
  m^2_{\chi}}\,[\bar{u}_{\chi}(\vec{k}_{\chi},
  \sigma_{\chi})\gamma^{\mu}(1 + \gamma^5) u_n(\vec{k}_n,
  \sigma_n)]\,\frac{\displaystyle - \eta_{\mu\nu} +
  \frac{q_{\mu}q_{\nu}}{M^2_{Z'}}}{M^2_{Z'} - q^2 -
  i0}\nonumber\\ &&\times\,[\bar{u}_{\ell}(\vec{k}_{\ell})
  \gamma^{\nu}(1 - \gamma^5)v_{\bar{\ell}}(\vec{k}_{\bar{\ell}},
  \sigma_{\bar{\ell}})].
\end{eqnarray}
Summing up the contributions of the Feynman diagrams in
Fig.\,\ref{fig:fig5} we obtain the amplitude of the neutron dark
matter decays $n \to \chi + \ell + \bar{\ell}$
\begin{eqnarray}\label{eq:42}
M(n \to \chi + \ell + \bar{\ell}) &=& - \frac{g_{\chi}e^2_{\chi}}{2
  M^2_{Z'}}\,\frac{m^2_{\chi}}{m^2_n -
  m^2_{\chi}}\,\Big[\bar{u}_{\chi}(\vec{k}_{\chi},
  \sigma_{\chi})\gamma^{\mu}(1 + \gamma^5) u_n(\vec{k}_n,
  \sigma_n)\Big]\,\frac{M^2_{Z'}}{M^2_{Z'} - q^2 - i0}\Big(-
\eta_{\mu\nu} + \frac{q_{\mu}q_{\nu}}{M^2_{\cal
    Z}}\Big)\nonumber\\ &&\times\,[\bar{u}_{\ell}(\vec{k}_{\ell})
  \gamma^{\nu}(1 - \gamma^5)v_{\bar{\ell}}(\vec{k}_{\bar{\ell}},
  \sigma_{\bar{\ell}})].
\end{eqnarray}
Assuming that $M^2_{Z'} \gg q^2$ we arrive at the amplitude
\begin{eqnarray}\label{eq:43}
M(n \to \chi + \ell + \bar{\ell}) = \frac{g_{\chi}e^2_{\chi}}{2
  M^2_{Z'}}\,\frac{m^2_{\chi}}{m^2_n -
  m^2_{\chi}}\,\Big[\bar{u}_{\chi}(\vec{k}_{\chi},
  \sigma_{\chi})\gamma^{\mu}(1 + \gamma^5) u_n(\vec{k}_n,
  \sigma_n)\Big]\,[\bar{u}_{\ell}(\vec{k}_{\ell}) \gamma^{\nu}(1 -
  \gamma^5)v_{\bar{\ell}}(\vec{k}_{\bar{\ell}}, \sigma_{\bar{\ell}})],
\end{eqnarray}
which can be obtained from the effective local Lagrangian
\begin{eqnarray}\label{eq:44}
{\cal L}_{\rm DMBL}(x) = \frac{g_{\chi}e^2_{\chi}}{2 M^2_{\cal
    Z}}\,\frac{m^2_{\chi}}{m^2_n -
  m^2_{\chi}}\,\big[\bar{\psi}_{\chi}(x)\gamma^{\mu}(1 +
  \gamma^5)\psi_n(x)\big]\,[\Psi_e(x)\gamma^{\nu}(1 - \gamma^5)
  \Psi_e(x)],
\end{eqnarray}
having the structure of the effective interaction Eq.(\ref{eq:5}),
where
\begin{eqnarray}\label{eq:45}
- \frac{G_F}{\sqrt{2}}V_{ud}h_V = \frac{g_{\chi}e^2_{\chi}}{2
  M^2_{Z'}}\frac{m^2_{\chi}}{m^2_n - m^2_{\chi}}\quad,\quad -
\frac{G_F}{\sqrt{2}}V_{ud}\bar{h}_A = \frac{g_{\chi}e^2_{\chi}}{2
  M^2_{Z'}}\frac{m^2_{\chi}}{m^2_n - m^2_{\chi}}.
\end{eqnarray}
In terms of the vacuum expectation value of the Higgs--field $v =
1/\sqrt{\sqrt{2}G_F} = 246\,{\rm GeV}$ the coupling constants $h_V$
and $\bar{h}_A$ are defined by
\begin{eqnarray}\label{eq:46}
h_V = \bar{h}_A = -
\frac{g_{\chi}}{2 V_{ud}}\,\frac{v^2}{v^2_{\chi}}\,\frac{m_{\chi}}{m_n -
  m_{\chi}}.
\end{eqnarray}
Taking into account the estimates Eq.(\ref{eq:13}) and $|g_{\chi}| <
2.45\times 10^{-3}\sqrt{m_n - m_{\chi}}/m_n$ we may estimate the
vacuum expectation value $v_{\chi}$. We get $v_{\chi} \sim 0.09\, v
\,(m_n - m_{\chi}) \sim 22\,(m_n - m_{\chi})\,{\rm GeV}$, where $m_n -
m_{\chi}$ is measured in MeV.  Below we extract the value of the mass
difference $m_n - m_{\chi}$ from the constraint on the suppression
scale of the dark matter production in the experiments of the ATLAS
Collaboration at the LHC.

\subsection{Suppression scale associated with effective interaction 
Eq.(\ref{eq:5})}

The effective interaction Eq.(\ref{eq:5}) or Eq.(\ref{eq:44}) we may
rewrite in terms of the suppression scale
\begin{eqnarray}\label{eq:47}
{\cal L}_{\rm DMBL}(x) &=& - \frac{1}{~~\Lambda^2_{n\chi}}
\big[\bar{\psi}_{\chi}(x)\gamma_{\mu} (1 + \gamma^5)\psi_n(x)\big]
    [\bar{\Psi}_e(x)\gamma^{\mu}(1 -
      \gamma^5)\Psi_e(x)],
\end{eqnarray}
where $\Lambda_{n\chi}$ is the suppression scale defined by
\begin{eqnarray}\label{eq:48}
\Lambda_{n \chi} = 2 v_{\chi} \sqrt{\frac{1}{|g_{\chi}|}\,\frac{m_n -
    m_{\chi}}{m_{\chi}}} > 890\,(m_n - m_{\chi})^{5/4}\,{\rm GeV},
\end{eqnarray}
where $m_n - m_{\chi}$ is measured in MeV. For $m_n - m_{\chi} \le
0.01\,{\rm MeV}$ we get $\Lambda_{n\chi} \sim 3\,{\rm GeV}$. For our
estimate $m_n - m_{\chi} \sim 0.12\,{\rm MeV}$, obtained from
comparison of the suppression scale with the constraints from the
experimental data by the ATLAS Collaboration at the LHC (see
discussion below Eq.(\ref{eq:51})), we get $\Lambda_{n\chi} \sim
63\,{\rm GeV}$.  Thus, the effective interaction Eq.(\ref{eq:5}) is
characterized by the suppression scale $\Lambda_{n\chi} \sim 63\,{\rm
  GeV}$ and the dark matter fermion mass $m_{\chi} < m_n \sim 1\,{\rm
  GeV}$, respectively.

\subsection{Comparison with experimental data on dark 
matter production in ATLAS experiments at the LHC}

The suppression scale Eq.(\ref{eq:48}) we cannot use for the
comparison with experimental data by the ATLAS Collaboration
\cite{ATLASReport1.2017,ATLASReport2.2017} at the LHC. Indeed, the
suppression scale Eq.(\ref{eq:48}) defines the strength of the $n \to
\chi$ transitions of the SM fermion into one dark matter fermion,
whereas in experiments by the ATLAS Collaboration it is assumed that
dark matter is produced in fermion--antifermion $\bar{\chi}\chi$
pairs.  Since in our model a mediator is a dark matter spin--1 boson
$Z'$ with mass $M_{Z'} \sim 22\,(m_n - m_{\chi})\, e_{\chi}\,{\rm
  GeV}$, we have to compare the predictions of our model with the
experimental constraints on the production of dark matter pairs
$\bar{\chi}\chi$, mediated by a spin--1 boson. According to
\cite{ATLASReport1.2017}, the theoretical model, which is used for the
analysis of experimental data by the ATLAS Collaboration, deals with a
spin--1 boson $Z'$ coupled to the $V + A$ dark matter
$\bar{\chi}\gamma^{\mu}(1 + \gamma^5)\chi$ and quark
$\bar{q}\gamma^{\mu}(1 + \gamma^5)q$ with coupling constants
$g_{\chi}$ and $g_q$, respectively \cite{ATLASModel1} (see Eqs.(2.1)
and (2.2) of Ref.\cite{ATLASModel1}). For the comparison of the
predictions of our model with the experimental data by the ATLAS
Collaboration we have to determine the amplitudes of $n\bar{n} \to
\chi\bar{\chi}$ and $n\bar{n} \to \ell\bar{\ell}$ annihilation
mediated by the spin--1 boson $Z'$.

\begin{figure}
\centering \includegraphics[height=0.19\textheight]{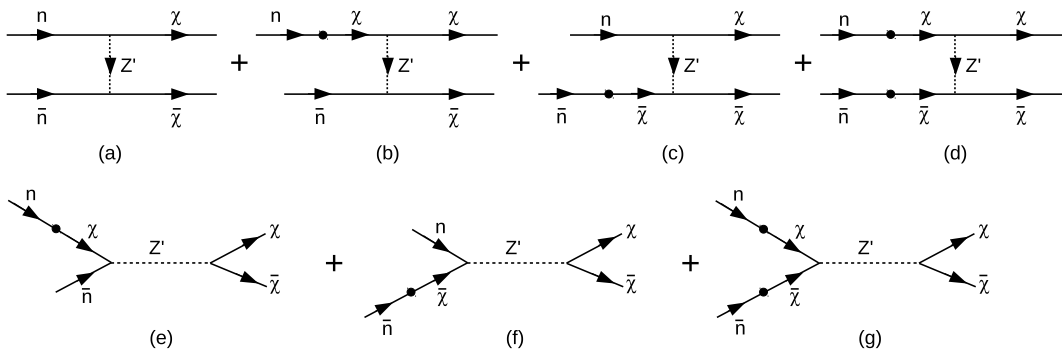}
  \caption{The Feynman diagrams, describing the amplitude of the
    process $n\bar{n} \to \chi\bar{\chi}$ mediated by the spin--1
    boson $Z'$.}
\label{fig:fig6}
\end{figure}

\begin{figure}
\centering \includegraphics[height=0.10\textheight]{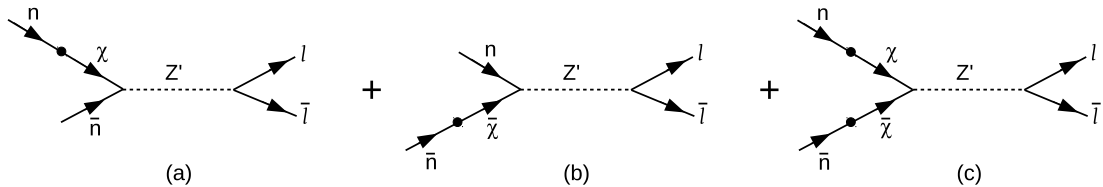}
  \caption{The Feynman diagrams, describing the amplitude of the
    process $n\bar{n} \to \ell\bar{\ell}$ mediated by the spin--1
    boson $Z'$.}
\label{fig:fig7}
\end{figure}

\begin{figure}
\centering \includegraphics[height=0.35\textheight]{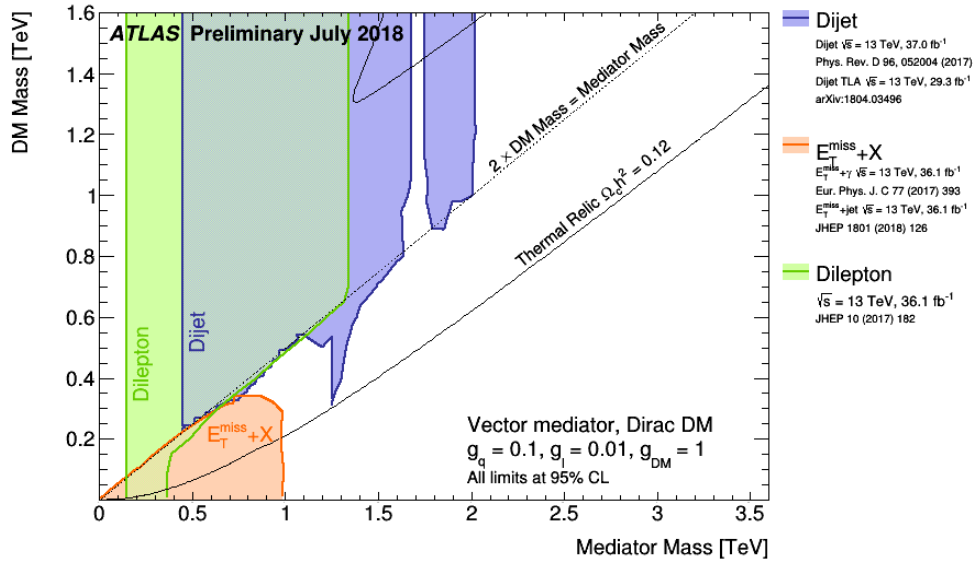}
  \caption{The experimental exclusion plots on the masses of the dark
    matter fermion independence of the mass of the dark matter spin--1
    boson by the ATLAS Collaboration \cite{ATLAS}.}
\label{fig:figAtlas}
\end{figure}

The amplitude of the $n\bar{n} \to\chi \bar{\chi}$ annihilation is
defined by the Feynman diagrams in Fig.\,\ref{fig:fig6}. The analytical
expression is equal to
\begin{eqnarray}\label{eq:49}
\hspace{-0.3in}M(n \bar{n} \to \chi \bar{\chi}) &=&
\frac{g^2_{\chi}e^2_{\chi}}{4M^2_{Z'}}\,\frac{m^2_n}{(m_n -
  m_{\chi})^2}\, [\bar{u}_{\chi}(\vec{k}\,'_1, \sigma'_1)
  \gamma^{\mu}(1 + \gamma^5) u_n(\vec{k}_1, \sigma_1)]\,\frac{
  M^2_{Z'}}{M^2_{Z'} - q^2 - i0}\Big(- \eta_{\mu\nu} +
\frac{q_{\mu}q_{\nu}}{4 M^2_{Z'}}\Big)\nonumber\\
\hspace{-0.3in}&&\times\,[\bar{v}_n(\vec{k}_2, \sigma_2)
  \gamma^{\nu}(1 + \gamma^5) v_{\chi}(\vec{k}\,'_2,
  \sigma'_2)]\nonumber\\
\hspace{-0.3in}&+&
\frac{g^2_{\chi}e^2_{\chi}}{4  M^2_{Z'}}\,\frac{m^2_n}{(m_n -
  m_{\chi})^2}\,[\bar{u}_{\chi}(\vec{k}\,'_1, \sigma'_1)
  \gamma^{\mu}(1 + \gamma^5) v_{\chi}(\vec{k}\,'_2,
  \sigma'_2)]\,\frac{M^2_{Z'}}{M^2_{Z'} - p^2 - i0}\Big(-
\eta_{\mu\nu} + \frac{p_{\mu}p_{\nu}}{M^2_{Z'}}\Big)\nonumber\\
\hspace{-0.3in}&&\times\,[\bar{v}_n(\vec{k}_2,
  \sigma_2) \gamma^{\nu}(1 + \gamma^5) u_n(\vec{k}_1,
  \sigma_1)],
\end{eqnarray}
where $q = k'_1 - k_1 = k_2 - k'_2$ and $p = k_1 + k_2 = k'_1 +
k'_2$. In turn, the amplitude of the reaction $n\bar{n} \to
\ell\bar{\ell}$ is defined by the Feynman diagrams in
Fig.\ref{fig:fig7}. The analytical expression is equal to
\begin{eqnarray}\label{eq:50}
\hspace{-0.3in}M(n \bar{n} \to \ell \bar{\ell}) &=&
\frac{g^2_{\chi}e^2_{\chi}}{4 M^2_{Z'}}\,\frac{m^2_n}{(m_n -
  m_{\chi})^2}\,[\bar{u}_e(\vec{k}\,'_1, \sigma'_1)
  \gamma^{\mu}v_e(\vec{k}\,'_2,
  \sigma'_2)]\,\frac{M^2_{Z'}}{M^2_{Z'} - p^2 - i0} \Big(-
\eta_{\mu\nu} + \frac{p_{\mu}p_{\nu}}{M^2_{Z'}}\Big)\nonumber\\
\hspace{-0.3in}&&\times \,[\bar{v}_n(\vec{k}_2, \sigma_2)
  \gamma^{\nu}(1 + \gamma^5) u_n(\vec{k}_1, \sigma_1)].
\end{eqnarray}
From Eq.(\ref{eq:49}) and Eq.(\ref{eq:50}) we define the suppression
scale $\Lambda_{\rm DM}$. We get
\begin{eqnarray}\label{eq:51}
\Lambda_{\rm DM} = \frac{2 M_{Z'}}{|g_{\chi}|e_{\chi}}\frac{m_n -
  m_{\chi}}{m_n} >  18\,(m_n - m_{\chi})^{3/2}\,{\rm TeV},
\end{eqnarray}
where $m_n - m_{\chi}$ is measured in MeV. For the derivation of
Eq.(\ref{eq:51}) we have used that $|g_{\chi}| < 2.45\times
10^{-3}\,\sqrt{m_n - m_{\chi}}/m_n$ and $v_{\chi} \sim 22\,(m_n -
m_{\chi})\,{\rm GeV}$. According to \cite{ATLAS2017}, the suppression
scale should not be smaller than $790\,{\rm GeV}$. Setting
$\Lambda_{\rm DM} = 790\,{\rm GeV}$ we get $m_n - m_{\chi} \sim
0.12\,{\rm MeV}$. Such an estimate of the mass difference agrees well
with our assumption that the production of the electron--positron pair
in the neutron dark matter decay $n \to \chi + e^- + e^+$ may be below
the reaction threshold. For $m_n - m_{\chi} \sim 0.12\,{\rm MeV}$ we
get $v_{\chi} \sim 3\,{\rm GeV}$ and $M_{Z'} \sim 3\,e_{\chi}\,{\rm
  GeV}$, respectively. Setting also $e_{\chi} = 1$ we obtain that our
model is characterized by the dark matter fermion mass $m_{\chi} <
m_n$ and the mass of the dark matter spin--1 boson $M_{Z'} \sim
3\,{\rm GeV}$, i.e. $M_{Z'} > m_{\chi}$. As we show below this value
for the mass of the dark matter spin--1 boson $Z'$ is confirmed by the
experimental constraints on the cross sections for low--energy dark
matter fermion--electron scattering.  Using the experimental data by
the ATLAS Collaboration \cite{ATLAS} (see Fig.\,\ref{fig:figAtlas})
one may see that the our model with the suppression scale
$\Lambda_{\rm DM} \sim 790\,{\rm GeV}$, the mass of the dark matter
spin--1 boson $M_{Z'} \sim 3\,{\rm GeV}$ and the mass of the dark
matter fermion $m_{\chi} < m_n \sim 1\,{\rm GeV}$ belongs to the
allowed region in the close vicinity of the origin (see also Fig.\,6
of Ref.\cite{ATLAS2017}).
The dark matter spin--1 boson $Z'$ is unstable under decays $Z' \to
\chi\bar{\chi}$, $Z' \to e^- e^+$, $Z' \to \nu_e \bar{\nu}_e$, $Z' \to
n\bar{\chi}$ and $Z' \to \bar{n}\chi$ with the total width equal to
\begin{eqnarray}\label{eq:52}
\Gamma_{Z'} = \frac{e^2_{\chi}}{12\pi}\,M_{Z'} +
\frac{e^2_{\chi}}{24\pi}\,M_{Z'}\Big(1 -
\frac{m^2_{\chi}}{M^2_{Z'}}\Big)\sqrt{1 - \frac{4
    m^2_{\chi}}{M^2_{Z'}}} +
\frac{g^2_{\chi}e^2_{\chi}}{12\pi}\,M_{Z'}\Big(1 -
\frac{m^2_{\chi}}{M^2_{Z'}}\Big)\sqrt{1 - \frac{4 m^2_{\chi}}{M^2_{Z'}}}.
\end{eqnarray}
In the last term we have neglected the difference between neutron and
dark matter fermion masses.  For $e_{\chi} = 1$, $M_{Z'} = 3\,{\rm
  GeV}$ and $m_{\chi} = 0.9394454\,{\rm GeV}$ we get $\Gamma_{Z'} =
108\,{\rm MeV}$. The lifetime of the dark matter spin--1 boson ${Z'}$
is $\tau_{Z'} = 6.1\times 10^{-24}\,{\rm s}$. The branching ratio
${\rm Br}(Z' \to e^-e^+) = 0.37$, calculated in our model at $e_{\chi}
= 1$ and $M_{Z'} = 3\,{\rm GeV}$, is in qualitative agreement with the
branching ratio ${\rm Br}(Z' \to e^-e^+) > 0.23$ of a dark matter
spin--1 boson $Z'$ with mass $M_{Z'} \le 3\,{\rm GeV}$, obtained in
\cite{Curtin2014} by taking into account the kinetic $ Z'$--mixing
with the electroweak $Z$--boson and photon at the tree--level (see
Fig.\,13\,(b) of Ref.\cite{Curtin2014}). In our model the dark matter
spin--1 boson $Z'$ couples directly to the leptons and dark matter
fermion $\chi$.

Making a replacement of the neutron--antineutron ($n\bar{n}$) pair by
the quark--antiquark $(q\bar{q})$ pair with a coupling constant
$g^2_q$ for $g_q = 0.25$ \cite{ATLAS2017} we may estimate the cross
sections for the reactions $q\bar{q} \to \chi \bar{\chi}$ and
$q\bar{q} \to \ell \bar{\ell}$ mediated by the dark matter spin--1
boson $Z'$. At $\sqrt{s} = 13\,{\rm TeV}$ and for $M_{Z'} = 3\,{\rm
  GeV}$ and $\Lambda_{\rm DM} = 790\,{\rm GeV}$ we get
$\sigma(q\bar{q}\to \chi\bar{\chi})_{Z'} \sim 2 \times 10^{-40}\,{\rm
  cm^2}$ and $\sigma(q\bar{q}\to \ell \bar{\ell})_{Z'} \sim
10^{-48}\,{\rm cm^2}$, respectively.

\subsection{Comparison with experimental data on low--energy dark 
matter fermion--electron scattering}

For the comparison of the predictions of our model for the dark matter
fermion--electron elastic scattering we may calculate the
corresponding cross section. Using the Lagrangian Eq.(\ref{eq:38}) and
the conditions of the $\chi + e^- \to \chi + e^-$ reaction
\cite{Essig2012a,Essig2012,Essig2017} we obtain the following cross
section
\begin{eqnarray}\label{eq:53}
\sigma(\chi e^- \to \chi e^-)_{Z'} =
\frac{e^4_{\chi}}{16\pi}\,\frac{m^2_e}{(M^2_{Z'} - (m_{\chi} -
  m_e)^2)^2}.
\end{eqnarray}
For $e_{\chi} = 1$, $M_{Z'} = 3\,{\rm GeV}$, $m_{\chi} =
939.4454\,{\rm MeV}$ and $m_e = 0.511\,{\rm MeV}$ we get $\sigma(\chi
e^- \to \chi e^-)_{Z'} < 3\times 10^{-38}\,{\rm cm^2}$. Our estimate
agrees well with the XENON10 data obtained at $90\,\%$ (C.L.) and
plotted in Fig.\,3 of Ref.\cite{Essig2017} for the dark matter fermion
form factor $F_{\rm DM} = 1$ and $m_{\chi} < 1\,{\rm GeV}$.

\subsection{Neutron lifetime anomaly and dark matter dynamics in  
neutron stars}

The influence of the dark matter fermion $\chi$, which can appear in
the final state of the neutron dark matter decays, on dynamics of
neutron stars has been investigated in
\cite{McKeen2018,Baym2018,Motta2018,Cline2018}. The main result is
that dark matter fermions in the equilibrium state with the SM matter
of neutron stars do not destroy the possibility for neutron stars to
reach the maximum mass of about $2\,M_{\odot}$ \cite{Demorest2010},
where $M_{\odot}$ is the mass of the Sun \cite{PDG2018}, only for
$m_{\chi} > 1.2\,{\rm GeV}$. In other words dark matter fermions with
masses $m_{\chi} < m_n$, which can be responsible for the solution of
the neutron lifetime anomaly, are ruled out. A certain possibility for
existence of dark matter fermions $\chi$ with masses $m_{\chi} < m_n$
may appear in case of existence of an repulsive interaction between
dark matter fermions mediated by a sufficiently light dark matter
spin--1 bosons, the Compton wavelength of which is larger than
inter-particle distances in neutron stars \cite{McKeen2018}. Such a
possibility for dark matter fermions from the neutron decays has been
realized in scenario by Cline and Cornell \cite{Cline2018} within
$U'(1)$ gauge quantum field theory model with dark matter fermions
$\chi$ coupled to a dark matter photon $A'$, which mass is constrained
by $m_{A'}/g' \leq (45 - 60)\,{\rm MeV}$, where $g'$ is a gauge
coupling constant or a dark matter ``charge'' of dark matter
fermions. According to \cite{Cline2018}, the ratio $m_{A'}/g' \leq (45
- 60)\,{\rm MeV}$ depends on the nuclear equation of state and has
been derived from the requirement for neutron stars to have masses
compatible with $2\,M_{\odot}$ \cite{Demorest2010}. In dependence of
the value of the gauge coupling constant $g'$ the mass of the dark
matter spin--1 boson $A'$ can satisfy inequalities $m_{A'} < 2 m_e$
and $m_{A'} > 2 m_e$. Since, according to Cline and Cornell
\cite{Cline2018}, the neutron lifetime anomaly is explained by a fully
invisible decay mode $n \to \chi + A'$, the mass of the dark matter
spin--1 boson $A'$ obeys the constraint $m_n - m_{\chi} > m_{A'}$.  A
small mixing with a photon may allow the dark matter spin--1 boson
$A'$ to decay either into $A' \to e^- + e^+$ and $A' \to 3\gamma$ for
$m_{A'} > 2m_e$ or into $A' \to 3\gamma$ for $m_{A'} < 2m_e$,
respectively. The experimental exclusion of the decay mode $n \to \chi
+ \gamma$ \cite{Tang2018} is satisfied by the suppression of this
decay mode with respect to the decay mode $n \to \chi + A'$,
i.e. $\Gamma(n \to \chi + \gamma)/\Gamma(n \to \chi + A') \ll 1$
\cite{Cline2018}. In the model \cite{Cline2018} the processes
$q\bar{q} \to \chi\bar{\chi}$ are mediated by the scalar dark matter
bosons with masses larger than $1.55\,{\rm TeV}$. This does not
contradict the constraints by the ATLAS experiments.

Since in our approach to the neutron lifetime anomaly the mass of the
dark matter fermion is smaller than the neutron mass $m_{\chi} < m_n$,
we have to accept the mechanism of the influence of dark matter
fermions on dynamics of neutron stars, allowing to have masses of
about $2\,M_{\odot}$, developed by Cline and Cornell
\cite{Cline2018}. For this aim we extend the symmetry of our model
from $SU_L(2) \times U_R(1) \times U'_R(1)$ to $SU_L(2) \times U_R(1)
\times U'_R(1) \times U''_L(1)$, where $U''_L(1)$ is a new dark matter
gauge group. The Lagrangian invariant under $U''_L(1)$ gauge
transformations takes the form
\begin{eqnarray}\label{eq:54}
{\cal L}_{\rm DM''} &=& \bar{\psi}_{\chi
  L}i\gamma^{\mu}(\partial_{\mu} + i
\tilde{e}_{\chi}\tilde{C}_{\mu})\psi_{\chi L} -
\frac{1}{4}\,\tilde{C}_{\mu\nu}\tilde{C}^{\mu\nu} + (\partial_{\mu} -
i \tilde{e}_{\chi}\tilde{C}_{\mu})\Phi^*(\partial_{\mu} + i
\tilde{e}_{\chi}\tilde{C}_{\mu})\tilde{\Phi} + \tilde{\kappa}^2
|\tilde{\Phi}|^2 - \tilde{\gamma} |\tilde{\Phi}|^4\nonumber\\ &-& 2
\tilde{f}_{\chi}\big(\bar{\psi}_{\chi R}\psi_{\chi L} \Phi
\tilde{\Phi}^* + \bar{\psi}_{\chi L}\psi_{\chi R} \Phi^*
\tilde{\Phi}\big),
\end{eqnarray}
where $\tilde{C}_{\mu\nu} = \partial_{\mu}\tilde{C}_{\nu} -
\partial_{\nu}\tilde{C}_{\mu}$ is the field strength tensor operator
of the dark matter spin--1 field $\tilde{C}_{\mu}$, $
\tilde{e}_{\chi}$ is the dark matter ``charge'' of the left--handed
dark matter fermions and the dark matter complex scalar field
$\tilde{\Phi}$. The last term in Eq.(\ref{eq:54}) is obtained from the
term $f_{\chi}\big(\bar{\psi}_{\chi R}\psi_{\chi L} \Phi +
\bar{\psi}_{\chi L}\psi_{\chi R} \Phi^*\big)$ in Eq.(\ref{eq:32}) by
the replacement
\begin{eqnarray}\label{e:55}
\sqrt{2}\,f_{\chi}\big(\bar{\psi}_{\chi R}\psi_{\chi L} \Phi +
\bar{\psi}_{\chi L}\psi_{\chi R} \Phi^*\big) \to 2
\tilde{f}_{\chi}\big(\bar{\psi}_{\chi R}\psi_{\chi L} \Phi
\tilde{\Phi}^* + \bar{\psi}_{\chi L}\psi_{\chi R} \Phi^*
\tilde{\Phi}\big).
\end{eqnarray}
This implies that the mass of the dark matter fermion $\chi$ appears
in the phase of spontaneously broken $U'_R(1)\times U''_L(1)$
symmetry. The Lagrangian Eq.(\ref{eq:54}) describes interactions of
dark matter particles only. We would like to notice that the SM
particles and the dark matter particles transforming under
$SU_L(2)\times U_R(1) \times U'_R(1)$ gauge transformations are
invariant under gauge transformations of the $U''_L(1)$ group.

Following Kibble \cite{Kibble1967} and repeating the procedure
expounded above, namely, assuming i) to replace $\tilde{\Phi}$ by
$\tilde{\Phi} = e^{i\,\tilde{\alpha}_{\chi}}(\tilde{v}_{\chi} +
\tilde{\sigma})/\sqrt{2}$, ii) to make a gauge transformation
$\psi_{\chi L} \to e^{\,i\,\tilde{\alpha}_{\chi}}\psi_{\chi L}$, and
iii) to introduce a new spin--1 boson field $Z''_{\mu} =
\tilde{C}_{\mu} + \partial_{\mu}
\tilde{\alpha}_{\chi}/\tilde{e}_{\chi}$, where $\tilde{v}_{\chi} =
\sqrt{\tilde{\kappa}^2/\tilde{\gamma}}$ is the vacuum expectation
value of the dark matter scalar field $\tilde{\Phi}$, we arrive at the
Lagrangian
\begin{eqnarray}\label{eq:56}
{\cal L}_{\rm DM''} &=& \bar{\psi}_{\chi
  L}i\gamma^{\mu}(\partial_{\mu} + i\tilde{e}_{\chi}
Z''_{\mu})\psi_{\chi L} - m_{\chi}\bar{\psi}_{\chi}\psi_{\chi}
- \frac{1}{4}\,Z''_{\mu\nu} Z''^{\mu\nu} + \frac{1}{2}\,M^2_{Z''}
Z''_{\mu} Z''^{\mu} + \tilde{e}^2_{\chi}\tilde{v}_{\chi} Z''_{\mu}
Z''^{\mu}\tilde{\sigma} + \frac{1}{2}\, Z''_{\mu} Z''^{\mu}
\tilde{\sigma}^2\nonumber\\
 &+&\frac{1}{2}\,\partial_{\mu}\tilde{\sigma} -
\frac{1}{2}\,m^2_{\tilde{\sigma}} \tilde{\sigma}^2 +
\tilde{\gamma}\,\tilde{v}_{\chi} \tilde{\sigma}^3 -
\frac{1}{4}\,\tilde{\gamma}\,\tilde{\sigma}^2 -
\tilde{f}_{\chi}\bar{\psi}_{\chi}\psi_{\chi}\Big(v_{\chi}\tilde{\sigma}
+ \tilde{v}_{\chi} \sigma + \sigma\tilde{\sigma}\big),
\end{eqnarray}
where $m_{\chi}$, $M_{Z''}$ and $m_{\tilde{\sigma}}$ are masses of the dark
matter fermion $\chi$, dark matter spin--1 $Z''$ and dark matter
scalar $\tilde{\sigma}$ fields
\begin{eqnarray}\label{eq:57}
m_{\chi} = \tilde{f}_{\chi}\,v_{\chi}\tilde{v}_{\chi} \quad,\quad
M_{Z''} = \tilde{e}_{\chi} \tilde{v}_{\chi} \quad,\quad
m_{\tilde{\sigma}} = \sqrt{2 \tilde{\gamma}}\, \tilde{v}_{\chi}.
\end{eqnarray}
Without loss of generality we may again set the mass of the dark
matter scalar boson $\tilde{\sigma}$ arbitrary heavy
\cite{GellMann1960,Weinberg1967,Gasiorowicz1969,Ivanov2018b}. This
leads to the decoupling of the dark matter scalar boson
$\tilde{\sigma}$ from the dark matter fermion $\chi$ and the dark
matter spin--1 boson $Z''$ in agreement with the Appelquist--Carazzone
decoupling theorem \cite{Appelquist1975}.

Since the dark matter spin--1 boson $Z'$ is too heavy to provide a
repulsion at large inter--particle distances in neutron stars, so the
contribution of its repulsion should be taken into account as some
corrections to the repulsion produced by the dark matter spin--1 boson
$Z''$. Indeed, following  McKeen {\it et al.}
  \cite{McKeen2018} (see also \cite{Cline2018}), the pressure and
  energy density of neutron stars (or the equation of state of neutron
  stars) should acquire the corrections (see Eq.(\ref{eq:11}) of
  Ref. \cite{McKeen2018} and Eq.(\ref{eq:9}) of Ref.\cite{Cline2018})
\begin{eqnarray}\label{eq:58}
\Delta P_{\chi} = \Delta \epsilon_{\chi} =
\frac{1}{2}\,\Big(\frac{\tilde{e}^2_{\chi}}{4 M^2_{Z''}} +
\frac{e^2_{\chi}}{4 M^2_{Z'}}\Big)\,n^2_{\chi} =
\frac{\tilde{e}^2_{\chi}}{8 M^2_{Z''}}\,\Big(1 +
\frac{\tilde{v}^2_{\chi}}{v^2_{\chi}}\Big)\,n^2_{\chi} =
\frac{\tilde{e}^2_{\chi}}{8 M^2_{Z''}}\,\big(1 +
R_{\chi}\big)\,n^2_{\chi}
\end{eqnarray}
caused by the contributions of the dark matter spin--1 bosons $Z''$
and $Z'$, respectively, where we have used $M_{Z''} = \tilde{e}_{\chi}
\tilde{v}_{\chi}$ and $M_{Z'} = e_{\chi} v_{\chi}$. Perturbative
contributions of the dark matter spin--1 boson $Z'$ imply that the
ratio $R_{\chi} = \tilde{v}^2_{\chi}/v^2_{\chi}$ obeys the constraint
$R_{\chi} \ll 1$. We estimate $R_{\chi}$ below
Eq.(\ref{eq:61}). Having neglected the contribution of the dark matter
spin--1 boson $Z'$ to the equation of state we may deal with the dark
matter spin--1 boson $Z''$ only.

Thus, the part of the total Lagrangian ${\cal L}_{\rm SM + DM'
  + DM''}$, which should be responsible for dark matter dynamics in
neutron stars, can be written in the following form
\begin{eqnarray}\label{eq:59}
{\cal L}_{\rm SM + DM' + DM''} = \bar{\psi}_{\chi}(i\gamma_{\mu} -
m_{\chi})\psi_{\chi} - \frac{1}{4}\,Z''_{\mu\nu} Z''^{\mu\nu} +
\frac{1}{2}\,M^2_{Z''} Z''_{\mu} Z''^{\mu} - \frac{1}{2}\,\tilde
     {e}_{\chi}\, \bar{\psi}_{\chi}\gamma_{\mu}(1 -
     \gamma^5)\psi_{\chi} Z''_{\mu} + \ldots,
\end{eqnarray}
where the ellipsis denotes the contributions of other kinetic and
interaction terms of the SM and dark matter particles of the model. In
the non--relativistic approximation the potential of the dark matter
spin--1 boson $Z''$ between two dark matter fermions $\chi$ is equal
to
\begin{eqnarray}\label{eq:60}
V_{Z''}(r) = \frac{\tilde{e}^2_{\chi}}{16\pi}\,\frac{\displaystyle
  e^{- M_{Z''}r}}{r}.
\end{eqnarray}
Since it coincides with the potential of the vector field with mass
$M_{Z''}$, describing a repulsive interaction between two fermions
with ``charges'' $\tilde{e}_{\chi}/2$ separated by a distance $r$, we
may apply it for the analysis of dark matter dynamics in neutron stars
in the scenario by Cline and Cornell \cite{Cline2018}. For a short
confirmation of a validity of our model for the analysis of dynamics
of neutron stars we may use the estimate by Cline and Cornell
\cite{Cline2018}. Indeed, according to Cline and Cornell
\cite{Cline2018}, a possibility for neutron stars with dark matter
fermions lighter than neutron and light dark matter spin--1 bosons in
the equilibrium with the SM particles to reach maximum masses
compatible with $2\,M_{\odot}$ places the constraint (see Eq.(12) of
Ref.\cite{Cline2018}). Since the correction to the
  equation of state (see Eq.(\ref{eq:58}), caused by repulsion between
  dark matter fermions with mass $m_{\chi} < m_n$, is fully defined by
  the dark matter spin--1 boson $Z''$, the inequality Eq.(12) of
  Ref.\cite{Cline2018}) should be saturated only by the dark matter
  spin--1 boson $Z"$. In our notations such a constraint reads
\begin{eqnarray}\label{eq:61}
\frac{2 M_{Z''}}{\tilde{e}_{\chi}} \lesssim (45 - 60)\,{\rm MeV}.
\end{eqnarray}
This allows to estimate the vacuum expectation value
$\tilde{v}_{\chi}$.  Substituting $M_{Z''} =
\tilde{e}_{\chi}\tilde{v}_{\chi}$ into Eq.(\ref{eq:60}) we get
$\tilde{v}_{\chi} \lesssim (23 - 30)\,{\rm MeV}$.  Using
$\tilde{v}_{\chi} \lesssim (23 - 30)\,{\rm MeV}$ and $v_{\chi} \simeq
3\,{\rm GeV}$ for the ratio $R_{\chi} = \tilde{v}^2_{\chi}/v^2_{\chi}$
we get the value $R_{\chi} < 10^{-4}$. Thus, the contribution of the
dark matter spin--1 boson $Z'$ to the equation of state of neutron
stars makes up of about $0.01\,\%$ with respect to the contribution of
the dark matter spin--1 boson $Z''$. This, confirms our assertion that
the contributions of the dark matter spin--1 boson $Z'$ can be taken
into account perturbatively when it is required.

A specific value of the $Z''$--boson mass depends on the value of the
gauge coupling constant $\tilde{e}_{\chi}$, which can be obtained from
a detailed analysis of the interference of dark matter into dynamics
of neutron stars. Of course, such an analysis, i) using the dark
matter fermion mass obeying the constraint $m_{\chi} - m_n \simeq
0.12\,{\rm MeV}$, ii) taking into account the neutron dark matter
decay mode $n \to \chi + \nu_e + \bar{\nu}_e$, where the
neutrino--antineutrino pair possesses a zero net chemical potential
\cite{Baym2018}, and equations of state \cite{Motta2018,Gandolfi2012},
goes beyond the scope of this paper.  We are planning to carry out
such an analysis in our forthcoming publications. Here we would like
only to notice that there is practically nothing that can prevent for
the dark matter spin--1 boson $Z''$ to have a mass as light as the
dark matter spin--1 boson $A'$, introduced by Cline and Cornell
\cite{Cline2018}.

\subsection{URCA processes}

URCA processes were introduced by Gamow and Sch\"onberg
\cite{Gamow1941} for cooling of stars. As has been pointed out by
Gamow and Sch\"onberg \cite{Gamow1941}:''At the very high temperatures
and densities which must exist in the interior of contracting stars
during the later stages of their evolution, one must expect a special
type of nuclear processes accompanied by the emission of a large
number of neutrinos.'' According to \cite{Friman1979}, the process
\begin{eqnarray}\label{eq:62}
n + n \to n + n + \nu_e + \bar{\nu}_e
\end{eqnarray}
can be accepted as an URCA process and be also responsible together
with other URCA processes for the neutron star cooling. In our model
for the solution of the neutron lifetime problem there are processes
\begin{eqnarray}\label{eq:63}
n \to \chi + \nu_e + \bar{\nu}_e \quad, \quad n + n \to \chi +
\chi\quad,\quad n + n \to \chi + \chi + \nu_e + \bar{\nu}_e\quad,\quad
\chi + \chi \to n + n,
\end{eqnarray}
which can be also treated as URCA processes and give certain
contributions to the neutron star cooling. We are planning to carry
out an analysis of an influence of these processes on the neutron star
cooling in our forthcoming publications.

\section{Discussion}
\label{sec:conclusion}

We have analysed the dark matter scenario for the explanation of the
neutron lifetime puzzle. Following Fornal and Grinstein
\cite{Fornal2018} we have accepted the hypothesis that the neutron can
be unstable under dark matter decays. 

However, as we have emphasized from the very beginning such a
hypothesis is not innocent and entails the necessity to revise our
knowledge concerning the value either the axial coupling constant
$\lambda$ \cite{Abele2008,Abele2013}--\cite{Brown2018} or the Fierz
interference term $b$. Indeed, according to the hypothesis of an
existence of the neutron dark matter decay modes $n \to \chi + {\rm
  anything}$ \cite{Fornal2018}, the SM should explain the value
$\tau_n = 888.0\,{\rm s}$ of the neutron lifetime, measured in the
beam experiments, but not the value $\tau_n = 879.6\,{\rm s}$,
measured in the bottle ones. In this case using the results obtained
in \cite{Ivanov2013} one may show that the neutron lifetime $\tau_n =
888.0\,{\rm s}$ can be fitted at $\lambda = - 1.2690$. Since such a
value of the axial coupling constant is ruled out by experiments
\cite{Abele2008,Abele2013}--\cite{Brown2018}, the required value of
the neutron lifetime can be obtained only beyond the SM in terms of
the Fierz interference term.  At $\lambda = - 1.2750$ compatible with
the experimental values measured in \cite{Abele2013}--\cite{Brown2018}
we get the Fierz interference term equal to $b = - 1.44\times
10^{-2}$.  This is the price for the acceptance of the neutron dark
matter decays, explaining the neutron lifetime anomaly.

In addition to the dark matter decay mode $n \to \chi + e^- + e^+$
with the decay dark matter fermion $\chi$ and electron--positron pair,
proposed by Fornal and Grinstein \cite{Fornal2018}, we have added a
new dark matter decay mode $n \to \chi + \nu_e + \bar{\nu}_e$ with the
electron--neutrino--antineutrino pair. Such a decay mode should
explain the neutron lifetime puzzle in case of an unobservability of
the electron--positron pair. According to experimental analyses of the
dark matter decay mode $n \to \chi + e^- + e^+$, carried out in
\cite{Sun2018}, the branching fraction of such a process is suppressed
at the level of about $10^{-4}$ at $90\,\%$ (C.L.)  for kinetic
energies of the electron--positron pairs $100\,{\rm keV} \le T_{-+} <
644\,{\rm keV}$. Of course, there is still room for observation of the
electron--positron pairs with kinetic energies $T_{-+} < 100\,{\rm
  keV}$ by the UCNA Collaboration \cite{Sun2018} and by the PERKEO
Collaboration \cite{Klopf2018} using the electron spectrometer PERKEO
II \cite{Abele2016}. Our theoretical energy, angular and invariant
mass distributions can be used for the analysis for experimental data
by the UCNA and PERKEO Collaborations.

An unobservability of the electron--positron pairs in the products of
the neutron decay should not mean that the neutron dark matter
coupling $n \to \chi + e^- + e^+$ does not exist. This may also mean
that the electron--positron pair production is below of the reaction
threshold. In other words if the mass of the dark matter fermion obeys
the constraint $m_{\chi} > m_n - 2 m_e$, the dark matter decay mode $n
\to \chi + e^- + e^+$ is suppressed.  Because of such a possibility
the discrepancy between the neutron lifetimes, measured in the bottle
and beam experiments, can be explained by the contribution of the
neutron dark matter decay $n \to \chi + \nu_e + \bar{\nu}_e$. Since in
our model the strength of the interactions $n \to \chi + e^- + e^+$
and $n \to \chi + \nu_e + \bar{\nu}_e$ is the same, one may try to
observe the ``inverse'' neutron dark matter decay $e^- + n \to \chi +
e^-$ or simply the low--energy electron--neutron scatting. We have
calculated the differential cross section for the low--energy
electron--neutron scattering $e^- + n \to \chi + e^-$. We have found
that it possesses the following properties: i) it is inversely
proportional to the velocity of incoming electrons, ii) it is
isotropic, and iii) for low--energy incoming electron there is
constant flux of outgoing electrons with a momentum $k'_e = \sqrt{(m_n
  - m_{\chi})(m_n - m_{\chi} + 2m_e)}$, which is much larger than
3--momentum of incoming electrons.  Thus, in spit of a sufficiently
small value we may argue that the differential cross section for the
reaction $e^- + n \to \chi + e^-$ can be very well distinguished above
the background defined by the differential cross section for the
reaction $e^- + n \to n + e^-$ caused by the electromagnetic
electron--neutron couplings \cite{Hofstadter1956,Yennie1957} and
\cite{Schwinger1948,Dalitz1951,Gerasimov1963}.

For the UV completion of our effective interaction Eq.(\ref{eq:5}) we
have proposed a quantum field theory model invariant under $SU_L(2)
\times U_R(1) \times U'_R(1) \times U''_L(1)$ gauge
transformations. The sector of SM particles, including neutron,
proton, electron, neutrino, photon, electroweak bosons and
Higgs--boson, is described by the Standard Electroweak Model (SEM)
with $SU_L(2) \times U'_R(1)$ gauge symmetry. The gauge group
$U'_R(1)$ is responsible for the UV completion of the effective
interaction Eq.(\ref{eq:5}). The interactions between neutrons, dark
matter fermions and leptons are mediated by the dark matter spin--1
boson $Z'$. We have shown that the constraint on the suppression scale
$\Lambda_{\rm DM} \ge 790\,{\rm GeV}$ reported by the ATLAS
Collaboration \cite{ATLAS2017} is fulfilled in our model for the mass
differences $m_n - m_{\chi} \ge 0.12\,{\rm MeV}$. The mass of the dark
matter spin--1 boson $Z'$, defined for such a mass difference, is
$M_{Z'} \sim 3\,{\rm MeV}$ at $e_{\chi} = 1$. Such a value of the dark
matter spin--1 boson $Z'$ is confirmed by agreement of predictions of
our model for the cross section of low--energy dark matter
fermion--electron scattering ($\chi + e^- \to \chi + e^-$)
\cite{Essig2017}. Our
model with $SU_L(2) \times U_R(1) \times U'_R(1) \times U''_L(1)$
gauge symmetry and the dark matter fermion mass $m_{\chi} < m_n \sim
1\,{\rm GeV}$ and the dark matter spin--1 boson $Z'$ mass $M_{Z'}
\lesssim 3\,{\rm GeV}$ is not excluded by the experimental constraints
by the ATLAS Collaboration \cite{ATLAS} (see Fig.\,\ref{fig:figAtlas}
in the close vicinity of the origin).

Following our quantum field theory model and its agreement with
constraints by the ATLAS experiments one may state that a possible
observation of the neutron dark matter decay mode $n \to \chi + e^- +
e^+$ with kinetic energies of the electron--positron pair $T_{-+} <
100\,{\rm keV}$ by the UCNA Collaboration \cite{Sun2018} and by the
PERKEO Collaboration \cite{Klopf2018} should predict the suppression
scale $\Lambda_{\rm DM} > 18\,{\rm TeV}$ for dark matter
fermion--antifermion pair production in experiments by the ATLAS
Collaboration at the LHC.

The dark matter sector described by the gauge group $U''_L(1)$ is
responsible for dark matter dynamics in neutron stars in the
equilibrium state with the SM environment. As has been pointed out in
\cite{McKeen2018,Motta2018,Baym2018,Cline2018} the existence of dark
matter fermions $\chi$ with mass $m_{\chi} < m_n$ demands an existence
of light dark matter spin--1 bosons required for a repulsion between
dark matter fermions $\chi$ modifying the questions of state of
neutron stars and allowing neutron stars to have masses of about $2
M_{\odot}$. Since the dark matter spin--1 boson $Z'$ is too heavy to
describe correctly a required repulsion, we have introduced the dark
matter sector with a gauge $U''_L(1)$ group, which defines
interactions of left--handed dark matter fermions $\chi$ and a light
dark matter spin--1 boson $Z''$. Indeed, as we have shown above (see
Eq.(\ref{eq:58}) and discussion below Eq.(\ref{eq:61})) the
contribution of the dark matter spin--1 boson $Z'$ to the pressure and
energy density of neutron stars makes up of about $0.01\,\%$ with
respect to the contribution of the dark matter spin--1 boson $Z''$
belonging to the dark matter sector with $U''_L(1)$ gauge symmetry.
Such a dark matter sector is constructed according to scenario
developed by Cline and Cornell \cite{Cline2018}. However, unlike the
model by Cline and Cornell \cite{Cline2018} the dark matter spin--1
boson does not couple to photon. As a rough confirmation of an
applicability of our model to the analysis of dynamics of neutron
stars in the presence dark matter fermions $\chi$ with mass $m_{\chi}
< m_n$ and dark matter spin--1 bosons $Z''$ we have used the estimate
$2M_{Z''}/\tilde{e}_{\chi} \lesssim (45 - 60)\,{\rm MeV}$, obtained by
Cline and Cornell \cite{Cline2018}. This has allowed us to estimate
the vacuum expectation value $\tilde{v}_{\chi} \lesssim (23 -
30)\,{\rm MeV}$, defining the mass of the dark matter spin--1 boson
$Z''$. We have noticed that the use of the constraint
$2M_{Z''}/\tilde{e}_{\chi} \lesssim (45 - 60)\,{\rm MeV}$ from the
scenario by Cline and Cornell \cite{Cline2018} can be accepted as a
rough confirmation of an applicability of our model to the analysis of
dark matter dynamics in neutron stars.  Of course, a detailed analysis
demanding i) the use of the dark matter fermion mass $m_{\chi} - m_n
\simeq 0.12\,{\rm MeV}$, ii) the account for the neutron dark matter
decay mode $n \to \chi + \nu_e + \bar{\nu}_e$, where the
neutrino--antineutrino pair possesses a zero net chemical potential
\cite{Baym2018}, and iii) the use of equations of state
\cite{Motta2018,Gandolfi2012}, we are planning to carry out in our
forthcoming publications. In addition to the possibility of our model
to interfere into dynamics of neutron stars allowing to reach masses
of about $2 M_{\odot}$ we have argued that the processes $n \to \chi +
\nu_e + \bar{\nu}_e$, $n + n \to \chi + \chi$, $n + n \to \chi + \chi
+ \nu_e + \bar{\nu}_e$ and $\chi + \chi \to n + n$, mediated by the
dark matter spin--1 boson $Z'$, can be treated as URCA processes
\cite{Gamow1941,Friman1979,Hansel1995} and give a certain contribution
to the neutron star cooling.

We would like to notice that the masses of the dark matter scalar
bosons $\sigma$ and $\tilde{\sigma}$ in the phase of spontaneously
broken $U'_R(1) \times U''_L(1)$ symmetry are practically
arbitrary. In order to diminish the number of dark matter particles,
which can be, in principle, observable in terrestrial laboratories, we
have deleted dark matter scalar bosons $\sigma$ and $\tilde{\sigma}$
from their interactions with SM particles, dark matter fermions and
dark matter spin--1 boson $Z'$ and $Z''$ by setting their masses
infinitely heavy in agreement with the Appelquist-Carazzone decoupling
theorem \cite{Appelquist1975}. Such a decoupling is similar also to
the decoupling of the $\sigma$--meson in the linear $\sigma$-- model
(L$\sigma$M) of strong low--energy hadronic interactions
\cite{GellMann1960,Weinberg1967,Gasiorowicz1969,Ivanov2018b}.

We would like also to emphasize that our quantum field theory model
with gauge $SU(2)\times U(1) \times U'_R(1) \times U''_L(1)$ symmetry
predicts $n \leftrightarrow \chi$ oscillations
\cite{Babu2015}. Practically the models by McKeen {\it et al.}
\cite{McKeen2018} and Cline and Cornell \cite{Cline2018}, using a
certain $n \chi$ mixing should predict the $n \leftrightarrow \chi$
oscillations.

We would like to notice that recently \cite{Ivanov2018f}, we have
analysed the contribution of the interaction Eq.(\ref{eq:5}) to the
electrodisintegration of the deuteron $e^- + d \to \chi + p + e^-$
into dark matter and proton close to threshold. We have proposed to
search for such a dark matter channel $e^- + d \to \chi + p + e^-$ in
coincidence experiments on the electrodisintegration of the deuteron
$e^- + d \to n + p + e^-$ into neutrons $n$ and protons close to
threshold with outgoing electrons, protons and neutrons in
coincidence. A missing of neutron signals should testify a detection
of dark matter fermions.

\subsection{Abler--Bell--Jackiw anomalies and
    violation of renormalizability of renormalizable gauge theories}

We would like to notice that practical applications
  of the dark matter sector with $U'_R(1)$ gauge symmetry to the
  analysis of different processes with SM and dark matter particles
  can be restricted by tree-- and one--loop approximations. In the
  one--loop approximation the dark matter sector with $U'_R(1)$ gauge
  symmetry, described by the Lagrangian Eq.(\ref{eq:32}) is fully
  renormalizable and gauge invariant.

\begin{figure}
\includegraphics[height=0.08\textheight]{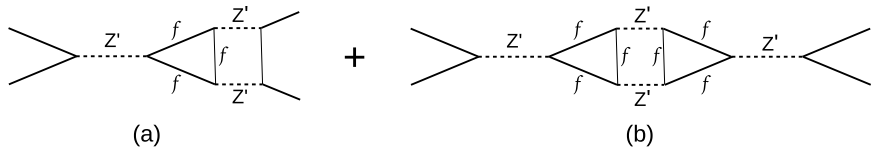}
  \caption{Examples of Feynman diagrams violating renormalizability of
    the dark matter sector with $U'_R(1)$ gauge symmetry to order
    $O(e^6_{\chi}/2^6)$ (a) and $O(e^8_{\chi}/2^8)$ (b) by the
    Adler--Bell--Jackiw anomaly in processes of fermion--fermion
    scattering or fermion--antifermion annihilation.}
\label{fig:fig9}
\end{figure}

Renormalizability of the dark matter sector with $U'_R(1)$ gauge
symmetry can be violated in higher order of perturbation theory
\cite{Bouchiat1972}--\cite{Bjorken1973} by the Adler--Bell--Jackiw
anomaly \cite{Adler1969,Bell1969}. A lowest order of perturbation
theory, to which renormalizability is violated by the
Adler--Bell--Jackiw anomaly, is $O(e^6_{\chi}/2^6)$, for example, in
the processes of fermion--fermion scattering or fermion--antifermion
annihilation. Some examples of Feynman diagrams of order
$O(e^6_{\chi}/2^6)$ and $O(e^8_{\chi}/2^8)$, violating
renormalizability of the amplitudes of fermion--fermion scattering or
fermion--antifermion annihilation by virtue of the Adler--Bell--Jackiw
anomaly in the dark matter sector with $U'_R(1)$ gauge symmetry, are
shown in Fig.\, \ref{fig:fig9} (see also Fig.\,14 of
Ref. \cite{Bjorken1973}).
The dark matter spin--1 boson $Z'$ couples to fermions through the
vertex $Z' Z' Z'$ described by one--fermion loops with virtual dark
matter fermions, electrons and neutrinos.  In order to restore
renormalizability to order $O(e^n_{\chi}/2^n)$, where $n \ge 6$, we
propose to add to the Lagrangian Eq.(\ref{eq:32}) the term
\begin{eqnarray}\label{eq:64}
\delta {\cal L}_{\rm DM'} = \bar{\psi}_{X
  R}i\gamma^{\mu}(\partial_{\mu} + i e_{\chi}C_{\mu})\psi_{X R} -
\sqrt{2}\, f_{X}\big(\bar{\psi}_{X R}\psi_{X L} \Phi + \bar{\psi}_{X
  L}\psi_{X R} \Phi^*\big) + \bar{\psi}_{X
  L}i\gamma^{\mu}\partial_{\mu}\psi_{XL},
\end{eqnarray}
where $\psi_{XR} = P_R\psi_X$ and $\psi_{XL} = P_L\psi_X$ are the
field operators of a dark matter fermion $X$. The Lagrangian
Eq.(\ref{eq:64}) is invariant under $U'_R(1)$ dark matter gauge
transformations 
\begin{eqnarray}\label{eq:65}
\psi_{X R} \to \psi'_{X R} = e^{\,i\alpha_{\chi}}\psi_{\chi
  R}\quad,\quad \Phi \to \Phi' = e^{\,i\alpha_{\chi}}\Phi \quad,\quad
\psi_{X L} \to \psi'_{X L} = \psi_{X L}\quad,\quad C_{\mu} \to C'_{\mu}
= C_{\mu} - \frac{1}{e_{\chi}}\,\partial_{\mu}\alpha_{\chi},
\end{eqnarray}
where $\alpha_{\chi}$ is a gauge parameter. In the physical phase the
Lagrangian Eq.(\ref{eq:65}) takes the form
\begin{eqnarray}\label{eq:66}
\delta {\cal L}_{\rm DM'} =
\bar{\psi}_X\big(i\gamma^{\mu}\partial_{\mu} - m_X)\psi_X -
\frac{1}{2}\,e_{\chi}\,\bar{\psi}_X\gamma^{\mu}(1 + \gamma^5)\,\psi_X
Z'_{\mu} - f_X \bar{\psi}_X\psi_X\sigma,
\end{eqnarray}
where $m_X = f_X v_{\chi}$ is a mass of the dark matter fermion $X$
such as $m_X = f_Xv_{\chi} \gg m_{\chi}$ and even $m_X \gg
M_{Z'}$. The anomalous diagrams are one--loop fermion $Z' Z'
Z'$--diagrams with a coupling constant $e^3_{\chi}/^3$. The
contributions of dark matter fermions $\chi$ and $X$ give the
Adler--Bell--Jackiw terms with a sign $(-1)$, whereas the electron and
neutrino contributions appear with the sign $(+1)$. Since the
Adler--Bell--Jackiw anomaly does not depend on the mass of virtual
fermions \cite{Adler1969,Bell1969}, the sum of the diagrams with dark
matter fermion $\chi$ and $X$, electron and neutrino loops is free
from the Adler--Bell--Jackiw anomaly.

\begin{figure}
\includegraphics[height=0.085\textheight]{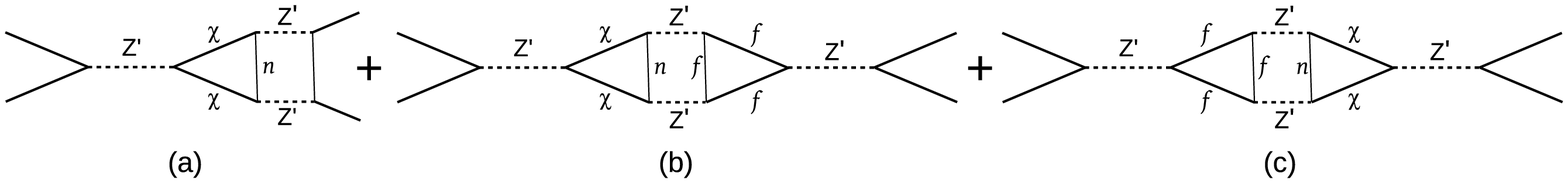}
  \caption{Example of Feynman diagrams violating renormalizability of
    the dark matter sector with $U'_R(1)$ gauge symmetry to order
    $O(g^2_{\chi}e^4_{\chi}/2^4)$ by the Adler--Bell--Jackiw anomaly,
    caused by the $n\chi Z'$ interaction.}
\label{fig:fig10}
\end{figure}

An additional violation of renormalizability by virtue of the
Adler--Bell--Jackiw anomaly can appear also because of the $n \chi Z'$
interaction.  For example, in the processes of fermion--fermion
scattering and fermion--antifermion annihilation the contribution of
the $n\chi Z'$ interaction, violating renormalizability by virtue of
the Adler--Bell--Jackiw anomaly, is of order $O(e^4_{\chi}
g^2_{\chi}/2^4)$ (see some examples of Feynman diagrams in
Fig.\,\ref{fig:fig10}).  Unfortunately, such a violation of
renormalizability cannot be repaired. It is important to emphasize
that a contribution of the Feynman diagrams, violating
renormalizability by virtue the Adler--Bell--Jackiw anomaly caused by
the $n \chi Z'$ interaction, relative to the main order contribution
$O(e^6_{\chi}/2^6)$ is of order $ O(4g^2_{\chi}/e^2_{\chi}) \sim
4\times 10^{-12}$ at $e_{\chi} = 1$ and $g_{\chi} < 2.45 \times
10^{-3}\,\sqrt{m_n - m_{\chi}}/m_n \sim 10^{-6}$ with $m_n - m_{\chi}
\simeq 0.12\,{\rm MeV}$. One may argue that violation of
renormalizability to such an order of perturbation theory with
contributions of a relative order $10^{-12}$ or even smaller cannot
discredit any quantum field theory model moreover when practical
applications of such a model to the analysis of observable phenomena
can be restricted by the tree-- and one--loop approximation only. This
is confirmed also by our analysis of the decay mode $H^0 \to Z + Z'$
of the SM Higgs--boson $H^0$ with mass $M_{H^0} = 125\,{\rm GeV}$,
where $Z$ is the electroweak boson.

According to \cite{Gross1972}, violation of renormalizability with a
relative order $10^{-12}$ or even smaller to order
$O(g^2_{\chi}e^4/2^4)$ and higher orders of perturbation theory,
caused by the $n \chi Z'$ interaction, may lead to violation of gauge
invariance only to the same order of magnitude, i.e. to a relative
order $10^{-12}$ or even smaller, and in the same orders of
perturbation theory. So we may argue that up to fourth order of
perturbation theory $O(g^2_{\chi}e^2_{\chi}/2^2) \sim 2\times
10^{-13}$ at $e_{\chi} = 1$ (see, for example, the Feynman diagram in
Fig\,\ref{fig:fig10}a without fermion line hooked by two dark matter
spin--1 boson $Z'$), describing fermion--dark matter spin--1 boson
$Z'$ scattering ($f + Z' \to f + Z')$ or fermion--antifermion
annihilation into $Z' Z'$--pair ($f + \bar{f} \to Z' + Z'$) and other
similar processes, renormalizability and gauge invariance of the dark
matter sector with $U'_R(1)$ gauge symmetry are not violated by the $n
\chi Z'$ interaction.

\subsection{Analysis of decay mode $H^0 \to Z
    + Z'$ of the Standard Model Higgs--boson $H^0$ with mass $M_{H^0}
    = 125\,{\rm GeV}$}

In addition to our estimates, which we have carried out in the main
part of our paper for confirmation of predictive power of our model,
we would like to analyse a compatibility of the values of the gauge
coupling constant $e_{\chi} = 1$ and the mass $M_{Z'} \sim 3\,{\rm
  GeV}$ of the dark matter spin--1 boson $Z'$, which we use in our
estimates, with recent analysis of exotic decay modes of the SM
Higgs--boson $H^0$ (or the Higgs--boson $H^0$) with mass $M_{H^0} =
125\,{\rm GeV}$, reported by Curtin {\it et al.}
\cite{Curtin2014}. For this aim we calculate the partial width of the
Higgs--boson decay mode $H^0 \to Z' + Z'$ and compare it with
constraints from Ref.\cite{Curtin2014}. The Higgs--boson $H^0$ with
mass $M_{H^0} = 125\,{\rm GeV}$, discovered by the ATLAS and CMS
Collaborations at the LHC \cite{ATLAS2012,CMS2012}, appears in the
phase of the spontaneously broken $SU_L(2) \times U_R(1)$ gauge
symmetry of the SM sector of our model.  The Lagrangian of the $H^0
ZZ$--interaction takes the form
\begin{eqnarray}\label{eq:67}
{\cal L}_{H^0 ZZ} = v\,\frac{1}{4}\,(g^2 + g'^2)\,H^0\,Z_{\mu}Z^{\mu}
= \frac{e^2}{4 \sin^2\theta_W \cos^2\theta_W}\,v\,H^0\,Z_{\mu}Z^{\mu},
\end{eqnarray}
where $H^0$ and $Z_{\mu}$ are the field operators of the Higgs--boson
$H^0$ with mass $M_{H^0} = 125\,{\rm GeV}$ \cite{ATLAS2012,CMS2012}
and the electroweak $Z$--boson, respectively \cite{PDG2018}. Then, $e
= \sqrt{4\pi \alpha}$ is the proton charge, expressed in terms of the
fine--structure constant $\alpha = 1/137.036$ \cite{PDG2018} and
related to the gauge coupling constants $g$ and $g'$ as follows $e = g
\sin\theta_W$ and $e = g'\cos\theta_W$, where $\theta_W$ is the
Weinberg angle $\sin^2\theta_W = 1 - M^2_W/M^2_Z = 0.223$ for $M_W =
80.379\,{\rm GeV}$ and $M_Z = 91.1876\,{\rm GeV}$, which are the
masses of the electroweak $W$-- and $Z$--bosons \cite{PDG2018}. The
Feynman diagrams of the decay mode $H^0 \to Z + Z'$ are shown in
Fig.\,\ref{fig:fig11}. The $Z \to Z'$ transition is defined in the
one--lepton loop approximation with electron and neutrino intermediate
states. Renormalization of the divergent contributions of the
one--lepton loops we carry out by using the procedure expounded in
\cite{Jegerlehner1991,Jegerlehner2011}. Skipping intermediate
calculations we give the partial width of the decay mode $H^0 \to Z +
Z'$:
\begin{figure}
\includegraphics[height=0.11\textheight]{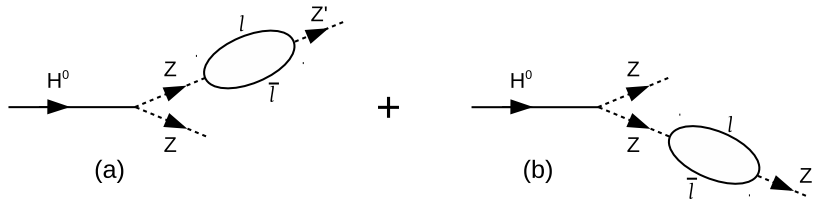}
  \caption{Feynman diagrams of the SM Higgs--boson decay mode $H^0 \to
    Z + Z'$. The $Z Z'$--mixing is caused by the one--lepton loop
    exchanges with electron and neutrino in the intermediate states.}
\label{fig:fig11}
\end{figure}
\begin{eqnarray}\label{eq:68}
\Gamma(H^0 \to Z Z') &=& M_{H^0}\,\frac{\alpha^3 e^2_{\chi}}{72\pi^2
  \sin^2\theta_W
  \cos^6\theta_W}\,\frac{v^2}{M^2_{H^0}}\,\frac{M^4_{Z'}}{(M^2_Z -
  M^2_{Z'})^2}\,\Big({\ell n}^2\frac{M^2_Z}{M^2_{Z'}} +
\pi^2\Big)\,\Big(1 + \frac{(M^2_{H^0} - M^2_Z - M^2_{Z'})^2}{8 M^2_Z
  M^2_{Z'}}\Big)\nonumber\\ &&\times \sqrt{\Big(1 - \frac{(M_Z +
    M_{Z'})^2}{M^2_{H^0}}\Big)\Big(1 - \frac{(M_Z -
    M_{Z'})^2}{M^2_{H^0}}\Big)} = 7.22\times 10^{-6}\,e^2_{\chi}\,{\rm
  MeV},
\end{eqnarray}
where $\pi^2$ is the contribution of the lepton pairs on--mass
shell. The numerical value is calculated for $M_{H^0} = 125\,{\rm
  GeV}$, $M_Z = 91\,{\rm GeV}$, $v = 246\,{\rm GeV}$ and $M_{Z'} =
3\,{\rm GeV}$.  Since the total width of the Higgs--boson is equal to
$\Gamma_{H^0} = 4.07\,{\rm MeV}$ \cite{PDG2018,Curtin2014}, the
branching ratio of the decay mode $H^0 \to Z + Z'$ is ${\rm Br}(H^0
\to Z Z') = 1.77\times 10^{-6}\,e^2_{\chi}$. For $e_{\chi} = 1$ the
branching ration ${\rm Br}(H^0 \to Z Z') = 1.77\times 10^{-6}$ agrees
well with constraints ${\rm Br}(H^0 \to ZZ') \sim 10^{-4} - 10^{-6}$
imposed by Curtin {\it et al.}  \cite{Curtin2014} (see Fig.\,12 of
Ref.\cite{Curtin2014}). Thus, the agreement of the predictions of our
model for the branching ratio ${\rm Br}(H^0 \to Z Z') = 1.77\times
10^{-6}$ with the results ${\rm Br}(H^0 \to Z Z') \sim (10^{-4} -
10^{-6})$, reported by Curtin {\it et al.}  \cite{Curtin2014}),
confirms fully correctness of the values of the gauge coupling
constant $e_{\chi} = 1$ and the mass $M_{Z'} \sim 3\,{\rm GeV}$ of the
dark matter spin--1 boson $Z'$, used in our model for numerical
analysis. A qualitative agreement of the branching ratio ${\rm Br}(Z'
\to e^-e^+) = 0.37$, calculated in our model, with the branching ratio
${\rm Br}(Z' \to e^-e^+) > 0.23$, proposed in \cite{Curtin2014}) (see
Fig.\,13/b) of Ref.\cite{Curtin2014}) does not suppress the use of the
parameters $e_{\chi} = 1$ and the mass $M_{Z'} \sim 3\,{\rm GeV}$.

\subsection{Formulation of our model with  $SU_L(2)  \times U_R(1) 
\times U'_R(1) \times U''_L(1)$ gauge symmetry at the quark level}

For the quark level analysis of the neutron lifetime anomaly and dark
matter production in ATLAS experiments at the LHC we propose to follow
\cite{Ioffe1981,Reinders1983,Ivanov1995,Ivanov1999} and present the
proton and neutron field operators in terms of three--quark densities
(the three--quark densities of other baryons from octet and decuplet
can be also found in
\cite{Ioffe1981,Reinders1983,Ivanov1995,Ivanov1999})
\begin{eqnarray}\label{eq:69}
\psi_p(x) \to \eta_p(x) &=&
\sqrt{\frac{2}{3}}\,\varepsilon^{ijk}[\bar{u^c}_i(x)
  \gamma_{\mu}u_j(x)]\,\gamma^{\mu}\gamma^5 d_k(x) -
\sqrt{\frac{1}{3}}\,\varepsilon^{ijk}[\bar{u^c}_i(x)
  \gamma_{\mu}d_j(x)]\,\gamma^{\mu}\gamma^5
u_k(x),\nonumber\\ \psi_n(x) \to \eta_n(x) &=&
\sqrt{\frac{1}{3}}\,\varepsilon^{ijk}[\bar{u^c}_i(x)
  \gamma_{\mu}d_j(x)]\,\gamma^{\mu}\gamma^5 d_k(x) +
\sqrt{\frac{2}{3}}\,\varepsilon^{ijk}[\bar{d^c}_i(x)
  \gamma_{\mu}d_j(x)]\,\gamma^{\mu}\gamma^5 u_k(x),
\end{eqnarray}
where $u_i(x)$ and $d_j(x)$ are the field operators of the {\it up}
and {\it down} quarks, respectively, $(i,j,k)$ are colour indices such
as $i(j,k) = 1,2,3$.Then, the quark field operator $\bar{q^c}_i(x)$ is
defined by $\bar{q^c}_i(x) = q^{T}_i(x)C$, where $C = - C^T = -
C^{\dagger}$ is the matrix of the charge conjugate, and $T$ is a
transposition.

At the quark level the SM sector of our model invariant under
$SU_L(2)\times U_R(1)$ gauge symmetry should have a standard form
including a complete set of left--handed quark doublets and
right--handed quark singlets and left--handed lepton doublets and
right--handed charged lepton singlets, respectively \cite{PDG2018}. In
turn, the dark matter sector invariant under $U''_L(1)$ gauge symmetry
is not changed, whereas the dark matter sector invariant under
$U'_R(1)$ gauge symmetry is described by the Lagrangian
\begin{eqnarray}\label{eq:70}
&&{\cal L}_{\rm DM'} = \bar{\psi}_{\chi R}i\gamma^{\mu}(\partial_{\mu}
  + i e_{\chi}C_{\mu})\psi_{\chi R} -
  \frac{1}{4}\,C_{\mu\nu}C^{\mu\nu} + (\partial_{\mu} -
  ie_{\chi}C_{\mu})\Phi^*(\partial_{\mu} + ie_{\chi}C_{\mu})\Phi +
  \kappa^2|\Phi|^2 - \gamma |\Phi|^4 \nonumber\\ &&+ \bar{\psi}_{\chi
    L}i\gamma^{\mu}\partial_{\mu}\psi_{\chi L} - \sqrt{2}\,
  f_{\chi}\big(\bar{\psi}_{\chi R}\psi_{\chi L} \Phi +
  \bar{\psi}_{\chi L}\psi_{\chi R} \Phi^*\big) +
  \bar{\Psi}_{eL}i\gamma^{\mu} \big(\ldots + ie_{\chi}
  C_{\mu}\big)\Psi_{eL} - 2 \zeta_e
  (\bar{\Psi}_{eL}\psi_{eR}\phi\,\Phi +
  \Phi^*\phi^{\dagger}\bar{\psi}_{eR}\Psi_{eL})\nonumber\\ &&+
  2\sqrt{2}\,\xi_{\chi}\,\big( \Phi^*\bar{\eta}_{n
    R}i\gamma^{\mu}(\partial_{\mu} + i e_{\chi}C_{\mu})\psi_{\chi R} -
  i(\partial_{\mu} - i e_{\chi}C_{\mu})\bar{\psi}_{\chi R}\gamma^{\mu}
  \eta_{n R} \Phi\big),
\end{eqnarray}
where $\eta_{nR} = P_R\eta_n$ is the three--quark field operator with
quantum numbers of the neutron and invariant under $U'_R(1)$ gauge
transformations. In the physical phase the Lagrangian, describing
interactions of dark matter with SM particles, takes the form
\begin{eqnarray}\label{eq:71}
{\cal L}_{\rm q\chi \ell} &=&
\xi_{\chi}\,v_{\chi}\,\big(\bar{\eta}_ni\gamma^{\mu}(1 +
\gamma^5)\partial_{\mu}\psi_{\chi} -
\partial_{\mu}\bar{\psi}_{\chi}i\gamma^{\mu}(1 + \gamma^5)\eta_n\big)
- \xi_{\chi} v_{\chi}\,e_{\chi}\big(\bar{\eta}_n\gamma^{\mu}(1 +
\gamma^5)\psi_{\chi} + \bar{\psi}_{\chi}\gamma^{\mu}(1 +
\gamma^5)\eta_n\big)Z'_{\mu}\nonumber\\ &-&
\frac{1}{2}\,e_{\chi}\bar{\psi}_{\chi}\gamma^{\mu}(1 + \gamma^5)
\psi_{\chi} Z'_{\mu} - \frac{1}{2}\,e_{\chi}
\bar{\Psi}_e\gamma^{\mu}(1 - \gamma^5) \Psi_e Z'_{\mu}.
\end{eqnarray}
The amplitude of the neutron dark matter decay mode $n \to \chi +
\ell + \bar{\ell}$ is defined by the Feynman diagrams in
Fig.\,\ref{fig:fig5} and by Eq.(\ref{eq:43}), where
$g_{\chi}u_n(\vec{k}_n,\sigma_n)$ is the matrix element
\begin{eqnarray}\label{eq:72}
g_{\chi}u_n(\vec{k}_n,\sigma_n) = \langle
0|\xi_{\chi}v_{\chi}\eta_n(0)|n(\vec{k}_n,\sigma_n)\rangle.
\end{eqnarray}
As a result, the effective low--energy Lagrangian of the neutron dark
matter decay modes is given by Eq.(\ref{eq:44}).

For the estimate of the suppression scale for the dark matter
production in the reaction $q \bar{q} \to \chi \bar{\chi}$ at
$\sqrt{s} = 13\,{\rm TeV}$ \cite{ATLAS2017} we may use the following
effective Lagrangian
\begin{eqnarray}\label{eq:73}
{\cal L}_{\eta_n \bar{\eta}_n \to \chi \bar{\chi}}(x) &\propto& -
\frac{\xi^2_{\chi} v^2_{\chi} e^2_{\chi}}{4
  M^2_{Z'}}\,[\bar{\psi}_{\chi}(x)\gamma^{\mu}(1 +
  \gamma^5)\,\eta_n(x)][\bar{\eta}_n(x) \gamma_{\mu}(1 +
  \gamma^5)\psi_{\chi}(x)] =\nonumber\\ &=& -
\frac{1}{4}\,\xi^2_{\chi}\,[\bar{\psi}_{\chi}(x)\gamma^{\mu}(1 +
  \gamma^5)\,\eta_n(x)][\bar{\eta}_n(x) \gamma_{\mu}(1 +
  \gamma^5)\psi_{\chi}(x)],
\end{eqnarray}
where $M_{Z'} = v_{\chi}e_{\chi}$. The effective Lagrangian
Eq.(\ref{eq:73}) is defined by the $t$--channel Feynman diagram in
Fig.\,\ref{fig:fig6}a. The contributions of the $t$-- and $s$--channel
Feynman diagrams in Fig.\,\ref{fig:fig6}b - Fig.\,\ref{fig:fig6}g can
be neglected for $\sqrt{s} = 13\,{\rm TeV}$ \cite{ATLAS2017}. Making a
Fierz transformation \cite{Fierz1937,Nieves2004} (see also
\cite{Itzykson1980}) we get
\begin{eqnarray}\label{eq:74}
{\cal L}_{\eta_n \bar{\eta}_n \to \chi \bar{\chi}} &\propto& -
\frac{1}{4}\,\xi^2_{\chi}\,[\bar{\eta}_n(x)\gamma^{\mu}(1 +
  \gamma^5)\,\eta_n(x)][\bar{\psi}_{\chi}(x) \gamma_{\mu}(1 +
  \gamma^5)\psi_{\chi}(x)].
\end{eqnarray}
The amplitude of the reaction $u_j \bar{u}^j \to \chi \bar{\chi}$,
where $u_j$ and $\bar{u}^j$ are the {\it up} quark and antiquark,
respectively, coupled at $\sqrt{s} = 13\,{\rm TeV}$ \cite{ATLAS2017},
is equal to
\begin{eqnarray}\label{eq:75}
M(u_j \bar{u}^j \to \chi \bar{\chi}) = \langle \chi\bar{\chi}|{\cal
  L}_{\eta_n \bar{\eta}_n \to \chi \bar{\chi}}(0)|u_j\bar{u}^j\rangle
\propto - \frac{1}{4}\,\xi^2_{\chi}\,[\bar{u}_{\chi}\gamma^{\mu}(1 +
  \gamma^5) v_{\chi}]\langle 0|[\bar{\eta}_n(0)\gamma^{\mu}(1 +
  \gamma^5)\,\eta_n(0)]|u_j\bar{u}^j\rangle.
\end{eqnarray}
Keeping only the leading contributions we obtain
\begin{eqnarray}\label{eq:76}
M(u_j \bar{u}^j \to \chi \bar{\chi}) \propto \frac{16}{3}\,\langle
\bar{d}d\rangle^2\,\xi^2_{\chi}\,[\bar{u}_{\chi}\gamma^{\mu}(1 +
  \gamma^5) v_{\chi}][\bar{v}^j\gamma^{\mu}(1 - \gamma^5)\,u_j],
\end{eqnarray}
where $v^j$ and $u_j$ are Dirac bispinor wave functions of the {\it
  up} antiquark and quark, respectively. The main contribution to the
amplitude of the reaction $u_j \bar{u}^j \to \chi \bar{\chi}$ comes
from the second term of the neutron quark structure
Eq.(\ref{eq:69}). The vacuum expectation value of the products of the
diquark and anti--diquark field operators can be expressed in terms of
the quark condensate $\langle \bar{d}d\rangle$. Since in the
$t$--channel transferred momenta are non--relativistic, for a rough
estimate of the suppression scale of the reaction $u_j \bar{u}^j \to
\chi \bar{\chi}$ we may use $\langle \bar{d}d\rangle = - (0.240\,{\rm
  GeV})^3$ determined at the scale of spontaneously broken chiral
symmetry \cite{Ioffe1981}. In this case the suppression
scale of the reaction $u_j \bar{u}^j \to \chi \bar{\chi}$ is given by
\begin{eqnarray}\label{eq:77}
\Lambda_{\rm DM} \sim  \frac{\sqrt[4]{3}}{4}\,\frac{1}{|\langle
  \bar{d}d\rangle||\xi_{\chi}|} \ge 790\,{\rm GeV},
\end{eqnarray}
where we have used the constraint $\Lambda_{\rm DM} \ge 790\,{\rm
  GeV}$ \cite{ATLAS2017}. This allows to estimate the coupling
constant $\xi_{\chi}$:
\begin{eqnarray}\label{eq:78}
|\xi_{\chi}| \sim \frac{\sqrt[4]{3}}{4}\,\frac{1}{|\langle
  \bar{d}d\rangle|\Lambda_{\rm DM}} = \frac{1}{\Lambda^4_{\chi}},
\end{eqnarray}
where $\Lambda_{\chi} \sim 2.4\,{\rm GeV}$.  Of course, we understand
that a consistent calculation of the amplitude of the reaction
$q\bar{q}\to \chi\bar{\chi}$ demands the use of gluon exchanges, and
our calculation of the amplitude of the reaction $u_j \bar{u}^j \to
\chi\bar{\chi}$ is sufficiently rough. Nevertheless, it can be used to
illustrate a way to estimate the suppression scale for the reactions
$q \bar{q} \to \chi\bar{\chi}$ investigated experimentally in
\cite{ATLAS2017}.  Thus, our analysis of the neutron lifetime anomaly
by means of the effective low--energy interaction Eq.(\ref{eq:5}) is
supported by the quantum field theory model with $SU_L(2)\times
U_R(1)\times U'_R(1) \times U''_L(1)$ gauge symmetry, which can be
formulated at the hadronic and quark level, respectively.

There are two problems, the solution of which goes beyond the scope of
this paper. They are i) an interference of dark matter into evolution
of neutron stars with dynamics described by the dark matter sector of
our model with $U''_L(1)$ gauge symmetry
\cite{McKeen2018}-\cite{Cline2018}(see also \cite{Karananas2018}) and
ii) a dark matter fermion--antifermion annihilation into lepton pairs
\cite{Galli2011}--\cite{D'Amico2018} (see also
\cite{Karananas2018}). The latter is due to i) a necessity to include
the dark matter fermion $X$ with mass $m_X \gg m_{\chi} \sim 1\,{\rm
  GeV}$ in order to avoid violation of renormalizability, caused by
the Adler--Bell--Jackiw anomalies, and ii) a possible asymmetry
between dark matter fermions and antifermions
\cite{Graesser2011}-\cite{Baldes2017}. We are planning to investigate
these problems within the framework of our model in our forthcoming
publications. Of course, the calculation of the amplitude of the
reaction $q \bar{q} \to \chi \bar{\chi}$ within the framework of our
model by taking into account gluon exchanges is also in the field of
our interest.

\section{Acknowledgements}

We are grateful to Hartmut Abele and Michael Klopf for numerous
interesting discussions, which were very important during the work
under this paper. The work of A. N. Ivanov was supported by the
Austrian ``Fonds zur F\"orderung der Wissenschaftlichen Forschung''
(FWF) under contracts P26781-N20 and P26636-N20 and ``Deutsche
F\"orderungsgemeinschaft'' (DFG) AB 128/5-2.  The work of
R. H\"ollwieser was supported by the Deutsche Forschungsgemeinschaft
in the SFB/TR 55.  The work of M. Wellenzohn was supported by the MA
23 (FH-Call 16) under the project ``Photonik - Stiftungsprofessur
f\"ur Lehre''.

\end{document}